# Pressure-induced superconductivity in weak topological insulator BiSe

Pallavi Malavi[a,1,2], Arpita Paul[b,1], Achintya Bera[a], D V S Muthu[a], Kunjalata Majhi[a], P S Anil Kumar[a], Umesh V. Waghmare[b], A. K. Sood[a,3] and S. Karmakar[c,3]

[a]Department of Physics, Indian Institute of Science, Bangalore, 560012, India;
[b]Theoretical Sciences Unit, Jawaharlal Nehru Centre for Advanced Scientific Research, Bangalore 560064, India;
[c]HP&SRPD, Bhabha Atomic Research Centre, Trombay, Mumbai 400085, India

## Abstract

Quasi-two-dimensional layered BiSe, a natural super-lattice with $Bi_2Se_3$-$Bi_2$-$Bi_2Se_3$ units, has recently been predicted to be a dual topological insulator, simultaneously weak topological insulator as well as topological crystalline insulator. Here using structural, transport, spectroscopic measurements and density functional theory calculations, we show that BiSe exhibits rich phase diagram with the emergence of superconductivity with Tc ~8K under pressure. Sequential structural transitions into SnSe-type energetically tangled orthorhombic and CsCl-type cubic structures having distinct superconducting properties are identified at 8 GPa and 13 GPa respectively. Our observation of weak-antilocalization in magneto-conductivity suggests that spin-orbit coupling (SOC) plays a significant role in retaining non-trivial band topology in the trigonal phase with possible realization of 2D topological superconductivity. Theoretical analysis reveals that SOC significantly enhances superconducting Tc of the high-pressure cubic phase through an increase in electron-phonon coupling strength. Simultaneous emergence of Dirac-like surface states suggests cubic BiSe as a suitable candidate for the 3D-topological superconductor.

[3]S. Karmakar    Email: sdak@barc.gov.in

**Introduction**

The search for topological superconductivity in various types of topological quantum materials has been of immense current research interest for possible exploration of Majorana quasiparticles in condensed matter physics and potential applications in fault-tolerant topological quantum computation [1,2]. Among these, 3D topological insulators (TIs) are bulk insulating quantum materials characterized by non-trivial Z2 band topology due to the presence of strong spin-orbit coupling (SOC) and time reversal symmetry (TRS) invariant, that exhibit spin-polarized gapless surface states (Dirac cone) on their boundaries. These are further classified as either strong (STI) or weak topological insulators (WTI) based on the TRS invariant criteria that protect the surface states. In a STI, TRS protects surface states on all surfaces (with an odd number of Dirac cones), whereas a WTI is considered as stacking of 2D TI layers and thus the surface states occur only on side surfaces (with an even number of Dirac cones) and the layer surface remains insulating [3-9]. On the other hand, topological crystalline insulators (TCI) are another class with non-trivial band topology where conducting surface states (with even number of Dirac cones) are protected by crystal mirror symmetry without the need for SOC [10,11]. A dual topological character has also been found in some 3D TIs where surface states can be controlled separately by either breaking TRS or crystal symmetry by an applied perturbation [12,13]. Most weak topological insulators are also found as suitable candidates for TCIs with corresponding gapless states on layered surfaces [8,13,14], making them ideal candidates for higher-order topological insulators [15,16].

Superconductivity in 3D TIs has been investigated by proximity effect, chemical doping as well as under pressure [17-21]. Cu-intercalated highly doped $Bi_2Se_3$ exhibits unconventional bulk superconductivity with Andreev bound states on its surface, characterizing topological superconducting state (TSC) [19, 20]. Efforts to induce superconductivity (SC) under high pressure to explore clean 3D topological superconductors without doping have led to structural transitions into topologically trivial metallic phases prior to the emergence of superconductivity [21-23]. Moreover, the 2D Majorana fermions existing on the surface of a 3D TSC are different from the non-Abelian Majorana fermions of a 2D TSC proposed for topological quantum computing [1, 2]. The possible realization of 2D TSC in weak topological insulators (WTIs) or higher order topological insulators (HOTIs) is thus highly sought after. Recently, quasi-1D bismuth halides [7, 24], and bismuth-bilayer sandwiched $Bi_2Se_3$, $Bi_2Te_3$ and BiTeI layered quasi-2D compounds are found to exhibit WTI and TCI states [25,13,8,15]. These compounds have so far been investigated to understand their topological nature and very few studies have been reported on their transport properties [24,25]. Although the transport measurements on the conducting side surfaces on these WTIs are challenging due to inaccessible cleavage side planes, studies on the dark (insulating) surface may reveal novel properties of the nontrivial bulk states. Pressure-induced superconductivity in WTI candidate materials has so far been reported in quasi-1D bismuth halides [24,26], where structure evolves into a highly disordered phase, possibly undergoing topological transitions to a trivial metallic state. A detailed high pressure investigations on BiSe, representing the other family of the WTIs, are thus of immense current interest.

**Results and Discussion**
**High pressure structures.** High pressure x-ray diffraction measurements at room temperature show series of structural phase transitions in the quasi-2D layered structured BiSe crystals (Fig. 1). Ambient trigonal structure (*P-3m1*, z=6) is found to be stable up to ~7.4 GPa. However, a closer look at the pressure variation of the trigonal c/a ratio shows a change in slope near ~3 GPa, indicating an iso-structural transition, further supported by the change in bulk compressibility and pressure variation of the Raman mode frequencies at this pressure, as discussed in *SI X-ray diffraction* and *SI Raman scattering*. At 7.8 GPa, the emergence of several new Bragg peaks signals a structural phase transition, with the trigonal structure persisting up to ~9 GPa. The new peaks can be indexed as a combination of two orthorhombic phases, *Cmcm* and *Pnma* (GeS-type), both persisting up to 18 GPa, the highest pressure of this measurement. Similar tangled high-pressure orthorhombic structures have earlier been reported for SnTe and SnSe compounds due to their closely matching enthalpies [27,28]. Interestingly, as followed by these compounds also, orthorhombic BiSe phases transform into CsCl-type cubic phase (*Pm-3m*) at pressures above 13 GPa. The structural parameters at high pressures were obtained by the Rietveld fit of the diffraction patterns up to 7.4 GPa, and the multi-phase Le-Bail profile fits at higher pressures (Fig. 1b-d). Detailed evolution of structural parameters of different BiSe phases under pressure is shown in Fig. S5 and Table S1. Figure 1e displays different BiSe lattices and phase transition sequence with increasing pressure; low pressure trigonal structure first undergoes a complete lattice re-arrangement by moving one set of bismuth atoms from the bilayer into the van der Waal gap between two $Bi_2Se_3$ quintuple layers to form the high pressure orthorhombic structure motif. An enhanced structural disorder in the high-pressure phases is apparent from the intrinsic broadened Bragg peaks compared to that of the low-pressure trigonal phase. The observed drastic change in the Raman spectra above 7.3 GPa, as shown in Fig. S6, can be associated with the structural transitions, corroborating the XRD results. The significant broadening of the Raman spectra can also be attributed to the reconstructive type phase transition of BiSe. Upon release of pressure, the cubic phase transforms back to the orthorhombic phase, and more interestingly, the orthorhombic phases reverse back to the ambient trigonal phase in this reconstructive phase transition (see Fig. 1a).

**First-principles calculations.** Calculated electronic structure of trigonal BiSe (*P-3m1* phase) reveals that it is metallic in nature if the spin-orbit coupling (SOC) is not included. Inclusion of SOC splits the doubly degenerate flat band at energy 150 meV above the Fermi level along the Γ-A direction causing band inversion and opens a gap at all k-vectors in the Brillouin zone (see Fig. S1). Theoretical analysis based on Fu and Kane's method [3] verifies BiSe in *P-3m1* structure as a weak topological insulator (Z2 invariant:(0;001)) with an indirect band gap of ~11 meV, as discussed in *SI Ab initio calculations*.

In theoretical analysis of pressure-induced successive structural transitions, the possible trigonal (*P-3m1*), orthorhombic (*Pnma* and *Cmcm*) and cubic (*Pm-3m*) structures of BiSe at each pressure have been optimized and the lattice parameters are found in good agreement with the XRD results. Figure 2a shows pressure variation of calculated enthalpies ∆H of orthorhombic and cubic phases with respect to trigonal phase of BiSe. ∆Hs of orthorhombic phases (*Cmcm* and *Pnma*) decrease monotonically and eventually become negative near 10 GPa, confirming the structural transition from trigonal phase to orthorhombic phase. As a result of having similar enthalpy, the *Cmcm* and *Pnma* phases may be kinetically stable near 10 GPa at room temperature, in agreement with experiments. The enthalpy variation also supports the stability of the cubic (*Pm-3m*) phase above 15 GPa. The pressure dependent variation in the estimated lattice parameters of *P-3m1* phase show an anomaly at 4.5 GPa and are in good agreement with XRD results (see Fig. S5). This anomaly is associated with the experimentally observed iso-structural transition near 3 GPa. The iso-structural transition is also evident in changes in pressure coefficients of calculated frequencies of Raman active modes in the *P-3m1* phase (see Fig. S8a), in agreement with high-pressure

Raman measurements. A detailed comparison of the calculated lattice parameters and phonon frequencies with experimental results is given in Tables S1 and S2.

Indirect band gap of trigonal *P-3m1* phase increases with pressure up to P=4 GPa. It starts decreasing and eventually vanishes above 9 GPa (Fig. 2b). Such reversal of the trend in P-dependence of the band gap above 4 GPa results from the change in conduction band minimum from Γ-K direction to L point (Fig. 2c and Fig. 2d). The indirect-indirect band gap crossover at this pressure is associated with the observed iso-structural transition. Our calculations show that the weak topologically insulating state in trigonal BiSe (with Z2 invariants (0;001)) remains unchanged up to 9 GPa and BiSe becomes metallic in all phases above 9 GPa.

To explore the possible non-trivial band topology of the high-pressure metallic phases of BiSe, electronic structures of symmetric (Se termination) and asymmetric (Bi and Se terminations) slab models for (001) surface of cubic BiSe (at 18 GPa) have been determined to investigate the surface states. In case of asymmetric slab, surface related states show linear band crossing near the M point of the Brillouin Zone, at 10meV below the Fermi level. Whereas, linear band crossing of surface states is observed near Γ point, at 180 meV above the Fermi level in the symmetric slab model, see *SI Surface State calculations* (Fig. S2 and S3). These results indicate possible Dirac semi-metallic surface states of the high-pressure cubic (*Pm-3m*) phase of BiSe.

**Emergence of superconductivity at high pressures.** Although BiSe has been predicted to be a weak topological insulator with a bulk band gap of ~11meV, single crystals are often grown as highly n-doped due to Se vacancies [25]. This results in shifting the Fermi level into the conduction band causing large bulk metallic conduction, similar to the strong TI $Bi_2Se_3$. Figure 3a shows temperature-dependent resistance R(T) plots at various pressures. The overall R(T) behavior changes only marginally up to 8.1 GPa, except at low temperatures. A resistance peak feature is observed below 8 K at 7.3 and 8.1 GPa (see Fig 3b). The magnetic field dependence and current variation measurements indicate this feature to be associated with the onset of the SC domains, see Fig S9, S10. The peak feature at the onset of the SC transition and also a two-step transition have been observed in high-pressure studies on many layered topological insulators [21,29,24,30] without much discussion. A detailed investigation on boron-doped diamond has revealed that the resistance peak or a two-step transition may originate due to the formation of Bosonic islands at the SC onset and subsequently achieving percolation threshold at a lower temperature initiating resistance drop [31,32]. This is attributed to the granular nature of the SC phase (intrinsic and extrinsic) into the non-SC phase. As the superconductivity in BiSe appears near the border of the structural transition, it is speculated that one or both of the high-pressure orthorhombic phases are superconducting. The signature of onset SC at 5.8 GPa in the trigonal phase is inferred from its distinct drop in resistance below 7K as well as its field dependence (see Fig. S10b). The above results are of tremendous importance from the point of view of the realization of the 2D topological superconducting state.

At 9.3 GPa, the resistance drops significantly in two steps at ~6.8K and ~2K (Fig. 3b). The broad hump-shaped resistance drop at further higher pressures can be attributed to the domain arrangement of the two coexisting SC phases (*Cmcm* and *Pnma*) having distinct SC Tc (as has also been confirmed by our theoretical calculations, to be discussed below). In a separate measurement (Run2), a two-step and bump-like feature is also noticed at 8.4 GPa with a reduced onset Tc (~2.6 K), see Fig. S11. Zero resistance below the SC transition at this pressure is achieved at 1.4K at low current excitation. At a higher pressure (13.8 GPa), R(T) curves change significantly. Although the overall metallic resistance decreases at this pressure, unconventional T-dependence (dR/dT<0) is observed below 150K. This can arise due to enhanced structural disorder, as evident from the broad XRD Bragg peaks and Raman modes. Note that the orthorhombic phases undergo a structural transition into CsCl-type cubic phase above 13 GPa. Our XRD results show that BiSe transformation into the cubic phase is nearly completed at 18 GPa. Therefore, the sharp SC transition near 8K at 18 GPa occurs in the CsCl-phase of BiSe. The systematic lowering of Tc with increasing magnetic field as well as with increasing current excitation confirm the SC transitions having distinct upper critical fields at different pressures, as shown in Fig. S12. The bulk nature of the observed SC under pressure is evident by the observed diamagnetic response in the magnetic susceptibility measurements [Fig. 3c]. To our surprise, upon complete release of pressure, SC persists with Tc ~2K, see Fig. S13. As BiSe structure reverses into the trigonal phase upon decompression, observed superconductivity in the pressure quenched sample may be assigned to the low pressure trigonal phase. Also, the broad XRD Bragg peaks in the release run indicate a highly disordered structure, supporting the unconventional T-dependence (dR/dT<0) over the entire T-range in its normal state. These results support the assignment of the onset SC at 5.8 GPa in the trigonal phase in the increasing pressure run. The persistence of SC in the pressure-quenched BiSe sample (by inducing structural disorder) provides a novel method for further investigations to explore topological superconductivity by surface probe techniques.

The pressure variation of superconductivity onset temperature (Tc) is summarized in Fig. 3d. The onset Tc shows a dome-shaped behavior in the orthorhombic phase, Tc reaching maximum (~7.5 K) at ~9 GPa above which it decreases gradually. With the emergence of the CsCl-type phase, Tc increases abruptly to ~8 K, and remains mostly unchanged with further increase of pressure. The distinct pressure variation of Tc and the SC transition width suggest their origins in different phases. At this point, it is noteworthy that for temperatures above Tc, no structural transition related anomaly is seen in R(T) at any pressure allowing us to conclude that the room temperature structure is maintained in the SC phases. The magnetic field dependence of R(T) curves at 9.3 GPa shows that the onset Tc decreases systematically with increasing magnetic field (see Fig S10). At a magnetic field of 7 T the SC transition almost smears out. The Tc-H plot, when fitted with the Ginzburg-Landau (GL) equation [$H_{c2} = H_{c2}(0)[(1-t^2)/(1+t^2)$ where t=T/Tc], estimates an upper critical field $H_{c2}(0)$ ~8.7T. This is well within the BCS weak-coupling Pauli limit of $H_p$=1.86*Tc(0)= 12.6 T. However, a quasi-linear $H_{c2}(T)$ can be seen within our measured T range. Such linear behavior observed in the $Bi_2Se_3$ SC state at high pressure has been ascribed to unconventional spin-orbit coupled superconductivity [21]. However, measurements at further low temperatures are needed for better understanding of the pair-breaking mechanism in the SC states in BiSe. Field-dependent measurements for the SC transitions at 18 GPa (CsCl-structured BiSe phase) estimate a much reduced upper critical field (~5 Tesla), see Fig. S12.

Electron-phonon coupling calculations confirm the superconductivity in the high pressure orthorhombic (*Pnma* and *Cmcm*) and cubic (*Pm-3m*) phases of BiSe. Figures 4a,b show the calculated vibrational spectra of BiSe in the *Cmcm* phase (without SOC) and *Pm-3m* phase (with and without SOC) at various pressures. The Eliashberg spectral functions of orthorhombic (*Cmcm* phase) and cubic BiSe indicate that the optical phonons at high symmetry points couple strongly with the electronic states. Hydrostatic pressure results in the hardening of phonon modes in both phases. The electron-phonon coupling (EPC) constant λ decreases monotonically with increasing pressure and is caused by hardening of phonons (Fig.4c). Consequently, the superconducting Tc (6

K at 12 GPa) of the *Cmcm* phase decreases with increasing pressure (Fig. 4d). The calculated Tc and its pressure dependence in the orthorhombic phase agree well with the experiment. We also estimate the EPC constant (0.7) and the Tc (4.2 K) in the orthorhombic *Pnma* phase at 12 GPa. The coexistence of two orthorhombic phases with dissimilar Tc explains the observed two-step SC transitions. To understand the observed significantly enhanced Tc as compared to the calculated Tc in the *Pm-3m* phase, we studied the effect of SOC on the electron-phonon coupling constant $\lambda$ and the Tc in this phase. SOC hardens the phonon modes (at high symmetry points) and enhances DOS at the Fermi level, which are expected to decrease the EPC constant ($\lambda$). SOC-induced enhancement of phonon linewidth is found to be responsible for the effective enhancement of $\lambda$ from 0.74 (without SOC) to 0.84, see *SI Effect of SOC on EPC*. This eventually leads to an enhancement of Tc from 4.9 K (without SOC) to 6.7 K (with SOC) at 18 GPa (Figs. 4c,d), in agreement with our experimental results. Enhancement of superconducting Tc due to SOC has earlier been predicted in Pb and $\alpha$-Pu [33,34]. The coexistence of Dirac like surface states (as discussed in previous section) and bulk superconducting state makes the cubic BiSe a suitable candidate for 3D-topological superconductor.

Our estimate of superconducting Tc in the metallic *P-3m1* phase at 10 GPa is about 0.3 K (without SOC). Our estimate of $\lambda$ (0.26 without SOC) in the *P-3m1* phase is at-least three times weaker than its estimates in other high-pressure phases (*Pm-3m*, *Pnma* and *Cmcm*). The observed higher value of onset Tc (~5K) at 6 GPa in the trigonal phase of BiSe can be attributed to strong SOC and the presence of intrinsic Se-vacancies that causes enhanced bulk conduction. The broad SC transition width and non-zero resistance can also be associated with the topology-related surface superconductivity in this phase, that calls for future investigations [35].

**Magnetoresistance under pressure.** The role of SOC on the bulk transport properties in the trigonal phase has been investigated by measuring in-plane magneto-transport under pressure on a freshly cleaved single crystal BiSe of thickness ~25 micron. Figure 5a shows the transverse magneto-resistance [MR=(R(B)-R(0))/R(0)] with the B field applied perpendicular to the measurement plane (*ab*-plane) at 2.5K at various pressures up to 5.8 GPa. The measurements were done at higher pressures at 8K (above Tc). As the measurements on the flake sample were done in the van der Pauw geometry, the asymmetry of the MR curves originates from the in-plane Hall contribution. The Hall contribution can be separated by symmetrizing the MR curves, MR$_{sym}$ (B)= [MR(B)+(MR(-B)]/2. The asymmetric part is the Hall contribution which is a small fraction of the actual Hall resistance R$_{xy}$(B) $\propto$ [MR(B)-MR(-B)]/2. At low temperature, the negative linear field dependent Hall resistance indicates that the sample is highly electron doped: carrier density n~3x10$^{20}$ /cm$^3$ and mobility $\mu$=R$_H$/$\rho_0$ ~260 cm$^2$/V.s, assuming single-type carrier. These are of the same order as previously reported on the bulk BiSe crystal, but significantly higher than the values in thin films [25]. The asymmetry in the MR data systematically reduces as P increases to 2 GPa, but at higher pressures, the asymmetry starts increasing in the opposite direction showing a sign reversal of the Hall contribution. This indicates a pressure-induced change in the electronic structure associated with the iso-structural transition in the trigonal BiSe near 3 GPa [upper inset of Fig. 5a]. The electronic structure modification at this pressure is further reflected in the measured in-plane resistance at room temperature with increasing pressure showing anomalous change in R(P) variation, see *SI Trigonal phase in-plane magneto-resistance* and Fig. S17a. The sign reversal of the Hall coefficient at ~3 GPa suggests the emergence of additional hole pockets with high mobility along with existing electron pockets. Also noticed at this pressure is the change in pressure-variation of the power-law exponent (n in $R = R_0 + AT^n$) for low temperature resistance and the field dependence of the transverse magneto-resistance [$MR = (\mu B)^\alpha$], see Fig. S17c-e. These results are in line with the results of band structure evolution near 4.5 GPa by first principle calculations (discussed above). The anomalous change in c/a ratio of the trigonal lattice and the change in pressure coefficient of Raman mode frequencies near this pressure suggest the possible formation of interlayer Se-Se bonding making the structure more of 3D character and resulting in reduced c-axis compressibility above ~3 GPa. Such structural modifications leading to iso-structural phase transitions are observed in layered chalcogenides [36]. The lower inset of Fig. 5a shows that MR$_{symm}$ marginally increases at higher pressures, but the power-law dependence (MR $\propto$ B$^{1.8}$, [37]) remains unchanged at all pressures. It is noteworthy that a somewhat linear magneto-resistance at 2.5K and at 5.8 GPa is not a result of structural inhomogeneity [38] or system reaching quantum limit [39], but due to resistance drop from the SC onset (Fig. S10b). We suggest that the irreversible nature of the iso-structural transition and associated electronic structural modification in the trigonal phase upon decompression may have resulted in the observed superconductivity in the pressure-quenched trigonal phase of BiSe.

Figure 5b shows in-plane resistance variation with magnetic field, R$_{xx}$(B), at various pressures. The cusp like enhanced positive MR at low field is due to weak anti-localization (WAL), the quantum correction to the carrier conductivity in bulk BiSe resulting from the strong SOC-induced Rashba spin splitting on the (001) plane [25,13]. The Rashba spin splitting in layered BiSe arises due to the electric field gradient as a result of charge transport from Bi$_2$ bilayer to Bi$_2$Se$_3$ QL [25]. The WAL feature in R(B) plots is observed up to 8.1 GPa, in the stability range of the trigonal phase. At 7.3 and 8.1GPa, the R(B) plots are shown for higher T (at 8K, above Tc), showing a weaker WAL feature. Figure 5c shows the magneto-conductivity at the lowest pressure (0.5 GPa) which has been fitted by the combination of normal metallic conductivity, $\sigma_M \sim \sigma_0 [1 + AB^2]^{-1}$ and the 2D WAL contribution $\sigma_{WAL,2D} \sim \ln B$ [40]. The WAL feature in an identical sample has also been studied at low temperatures in varying field orientations to confirm the 2D transport in this bulk crystal (see Fig S15). The inset of Figure 5c shows the negative magneto-conductivity of the WAL part, and a fitting with the standard Hikami-Larkin-Nagaoka equation for the 2D transport [41] gives parameters L$_\phi$ ~270nm and $\alpha$ ~75. This large value of $\alpha$ in highly n-doped BiSe bulk crystal suggests presence of many coherent 2D transport channels, as also seen for highly doped Bi$_2$Se$_3$ crystals [40,42]. The presence of WAL feature with similar fitting parameters throughout the pressure range of the trigonal phase clearly indicates that the 2D coherent carrier transport behavior is maintained up to 8 GPa (see Fig S16), even when the system undergoes an iso-structural electronic change at ~3 GPa. This is in line with the theoretical calculations showing non-trivial topology maintaining in the stability range of the trigonal phase.

**Summary**

In conclusion, WTI candidate BiSe has been studied for structural and transport properties under high pressures. X-ray diffraction measurements and first principles DFT calculations reveal a structural reconstruction of the trigonal BiSe lattice above 8 GPa into a SnSe-type orthorhombic structure with energetically tangled mixed phases. At further higher pressures above 13 GPa, CsCl-type cubic phase of BiSe is stabilized. Superconductivity with a transition temperature Tc ~7 K emerges near the boundary of the first structural transition, with a significant resistance drop noticed at 5.8 GPa in the trigonal structure. Structural granularity drives the SC state to evolve in an unusual fashion across the structural transitions. The transition temperature Tc shows a dome-shaped pressure-dependence in the intermediate orthorhombic phase, whereas it remains pressure-independent in the cubic phase. Spin-orbit coupling notably enhances the SC Tc of high-pressure cubic phase through electron-phonon coupling induced increment of phonon linewidth. First-principles calculations on the (001) surface of high-pressure cubic phase of BiSe predict Dirac like linear crossing of the surface-related bands. The observed superconductivity in the high-pressure phases of BiSe and the predicted Dirac like surface states in the cubic phase make the high-pressure BiSe phase a remarkable candidate for a 3D-topological superconductor. Magneto-conductivity measurements reveal weak anti-localization (WAL) feature showing coherent 2D transport of the bulk carrier channels. The WAL feature (experimental observation) and the non-trivial topology (theoretical prediction) remain unchanged in the stability range of the trigonal phase. The presence of strong SOC, as evident from the observed WAL feature in MR, thus plays a vital role in the emergence of superconductivity (with an enhanced Tc) within the WTI state at elevated pressure as well as upon pressure release. The observed onset of SC in the trigonal phase may thus be topological surface-related and indicate the possible emergence of 2D topological superconductivity. An iso-structural transition has been identified in trigonal BiSe near 3 GPa beyond which the structural modification causes pressure-induced bulk gap closing, preceding the SC transition.

**Materials and Methods**

Single crystal BiSe samples were grown using modified Bridgman method. DAC-based transport measurements were carried out via four-probe technique using a bath cryostat (Cryovac) and magnetic measurements using a 7T SQUID magnetometer (Cryogenics Ltd). First-principles calculations were performed within the generalized gradient approximation (GGA) with the Perdew-Burke-Ernzerhof (PBE) formula. See *SI Materials and Methods* for details.

**ACKNOWLEDGMENTS**
AKS thanks Nanomission Council and the Year of Science professorship of DST for financial support. XRD measurements were performed at Elettra synchrotron, Trieste under the proposal id 20195291. PM thanks DST-INSPIRE program for financial support. KM thanks CSIR for financial support. UVW acknowledges the support from a JC Bose National fellowship of SERB-DST. AP is thankful to the National Supercomputing Mission, JNCASR for providing computational resources.

**Author Contributions:** S.K., P.M. and A.K.S designed research; P.M., S.K., A.B., D.V.S.M., K.M., P.S.A.K. performed research (experiments); A.P. and U.W. performed research (first-principles calculations); P.M., S.K., and A.K.S. wrote the paper with inputs from all the authors.

**Keywords:** Topological insulator, high pressure, structural transition, superconductivity, weak-antilocalization

[1]These authors contributed equally to this work.
[2]Present address: HP&SRPD, Bhabha Atomic Research Centre, Mumbai, 400085, India.
P.M. performed research during her visit to HP&SRPD, Bhabha Atomic Research Centre.

**Figures and Tables**

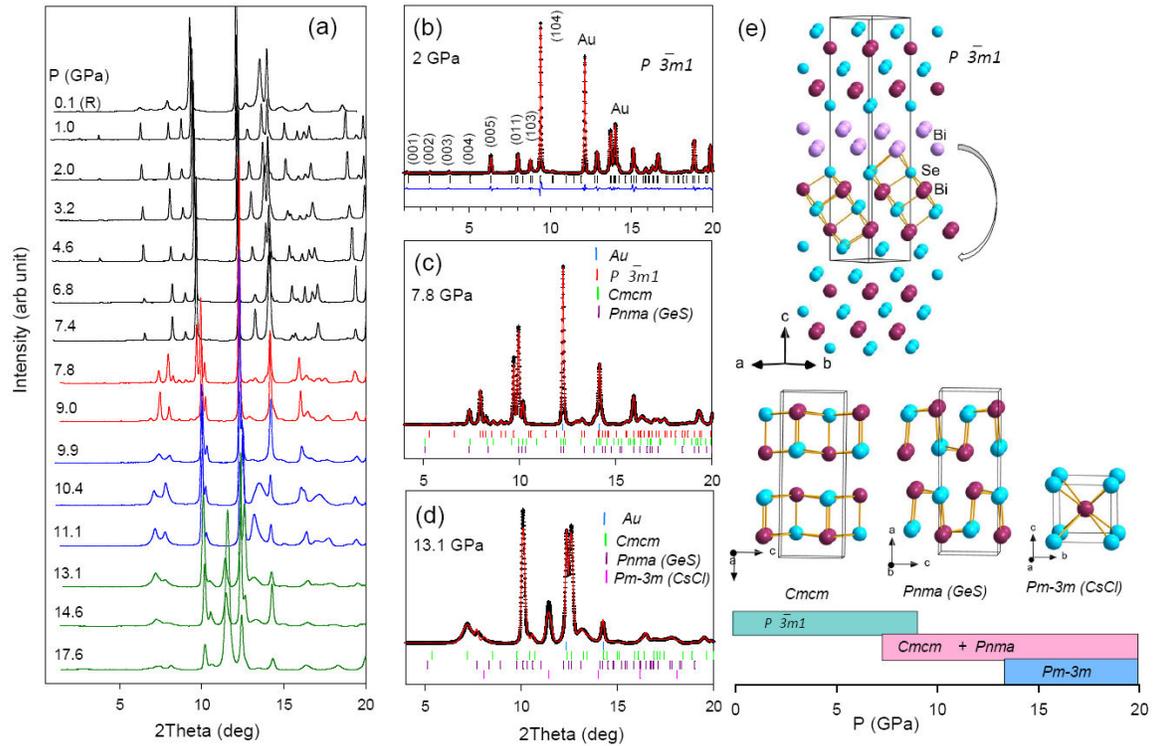

**Figure 1.** (a) X-ray diffraction patterns of BiSe at various high pressures (intensity arbitrarily shifted); black curves for trigonal phase, red curves for orthorhombic phases mixed with trigonal phase, blue curves for orthorhombic phases (with no trigonal phase) and green curves where CsCl-type structure emerges within mixed orthorhombic phases. The topmost pattern is at 0.1 GPa, taken from in the released run indicated by 'R'. Rietveld fits of the profiles (b) at 2 GPa (*P-3m1*), (c) at 7.8 GPa mixed phases, *Cmcm*, *Pnma* (GeS-type) and *P-3m1,* (d) at 13.1 GPa mixed phases, *Cmcm* and *Pnma* (GeS-type), and *Pm-3m* phase (CsCl type). (e) Top: Trigonal (*P-3m1*) structure of BiSe having $Bi_2Se_3$-$Bi_2$-$Bi_2Se_3$ unit cell (viewing direction [110]). Arrow indicates the shifting of position of one Bi layer into the Se-Se van der Waal layer causing total lattice reconstruction into a motif of smaller unit orthorhombic cell at 7.8 GPa. Middle: In the orthorhombic structures, *Cmcm* and *Pnma,* b- and a-axes respectively are oriented at 45° with respect to the trigonal c-axis. CsCl-type cubic structure can be viewed as smaller unit in the compressed orthorhombic cell. Bottom: The phase transition sequence of BiSe at room temperature with increasing pressure.

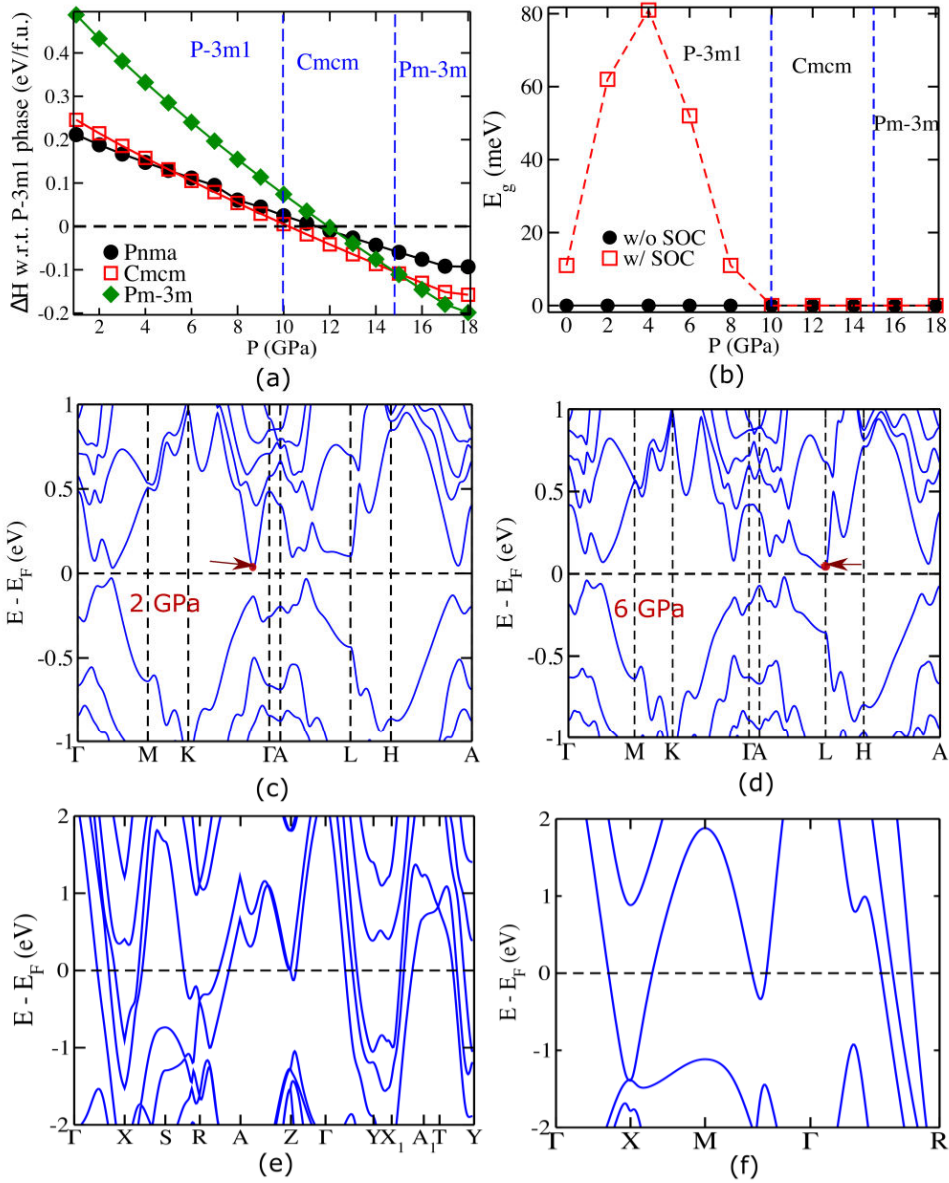

**Figure 2.** (a) Pressure dependent enthalpy (per formula unit) ΔH of orthorhombic *Pnma*, *Cmcm* and cubic *Pm-3m* phases with respect to *P-3m1* phase of BiSe, (b) Band gap of *P-3m1* phase as a function of pressure calculated with and without inclusion of SOC. Electronic structure (calculated including spin-orbit coupling) of *P-3m1* phase of BiSe at (c) P=2 GPa and (d) P=6 GPa. Changes in the positions of the conduction band minima, marked by red arrows in (c)-(d) explain the anomalous pressure dependence of band gap in (b). Electronic structure (including SOC) of (e) *Cmcm* (P=14 GPa) and (f) *Pm-3m* phases (P=18 GPa) of BiSe.

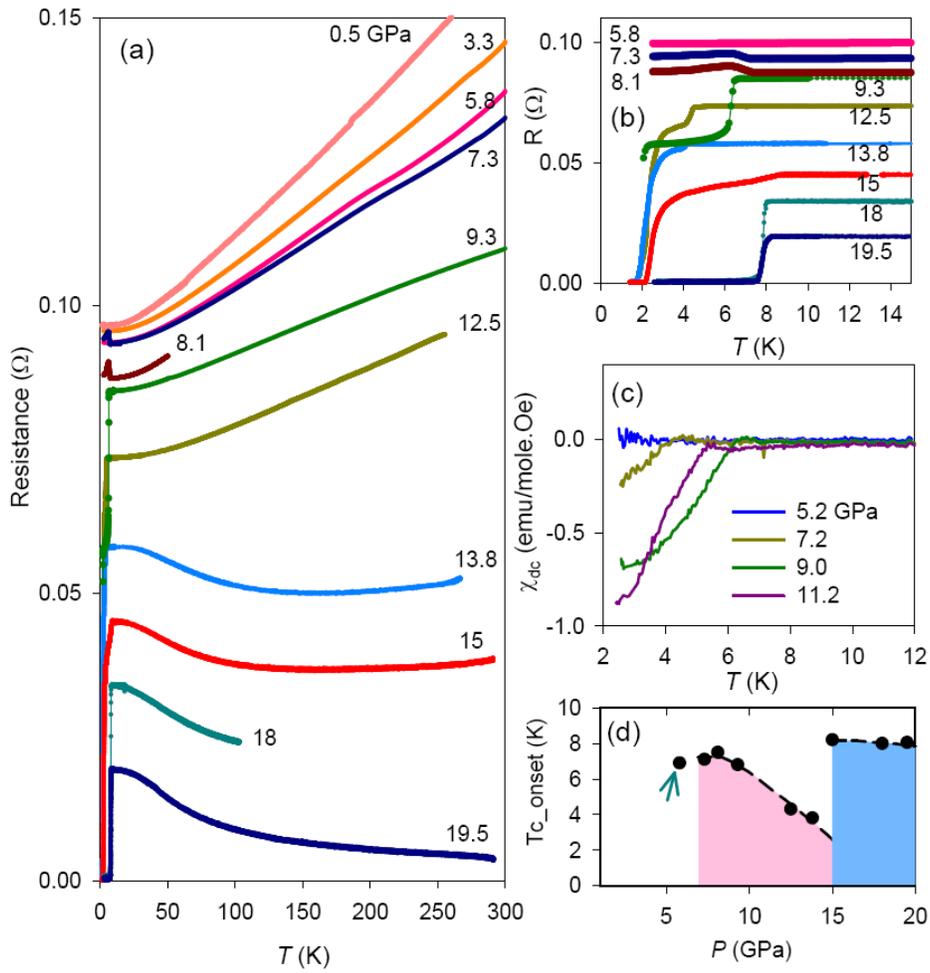

**Figure 3.** (a) Resistance of BiSe plotted as a function of temperature at various quasi-hydrostatic pressures. Resistance is measured along the ab-plane in the trigonal phase, (b) R(T) plots at low temperature near SC transition for pressures above 5.8 GPa. (c) ZFC dc susceptibility data at 50 Oe field from the sample at various high pressures. (d) Variation of SC onset Tc as a function of pressure. Arrow indicates the emergence of SC well within the stability range of trigonal (P-3m1) phase. Various SC regions are shaded according to Fig. 1e.

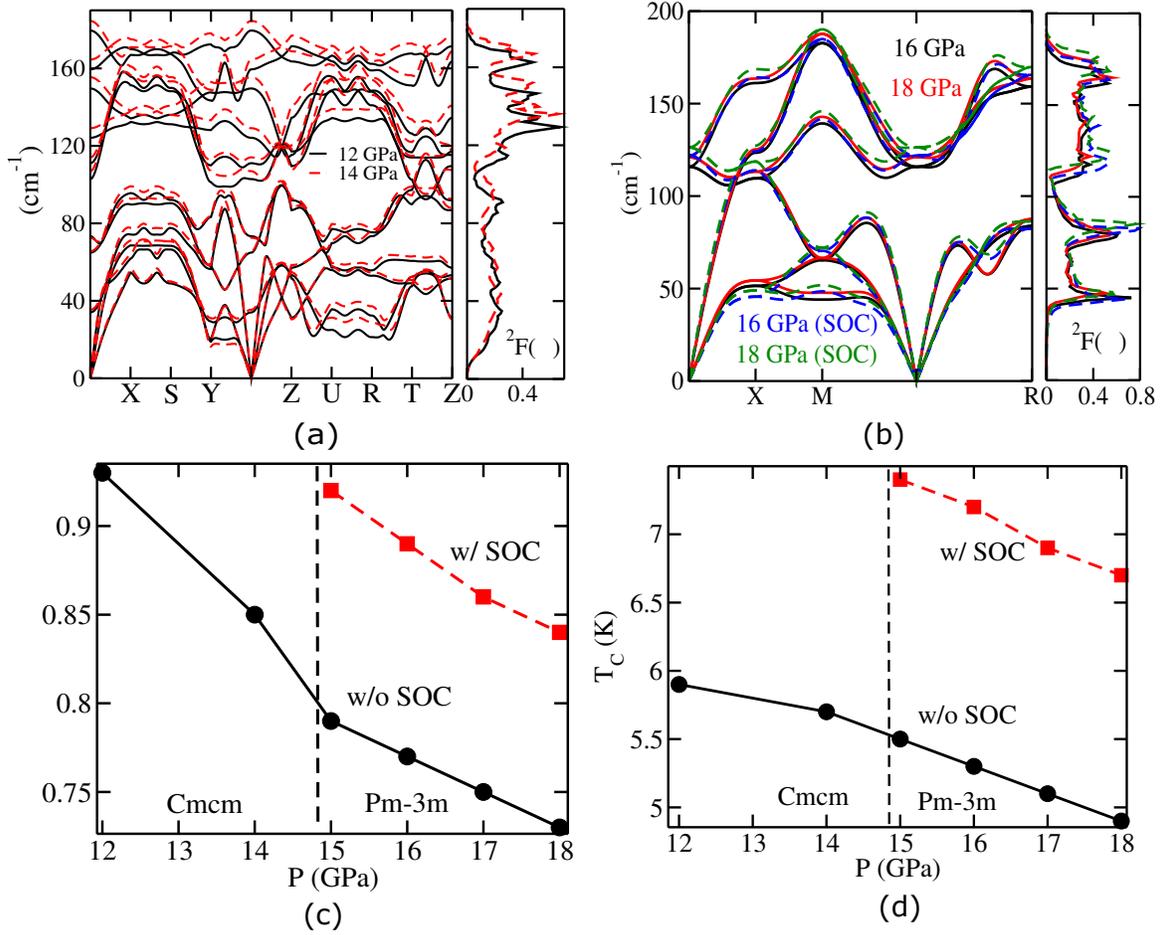

**Figure 4.** Pressure dependent phonon dispersion and Eliashberg spectral function in the (a) *Cmcm* phase (calculated without SOC), (b) *Pm-3m* phase (calculated with and without SOC) of BiSe, (c) electron phonon coupling constant $\lambda$ and (d) estimated superconducting transition temperature (Tc).

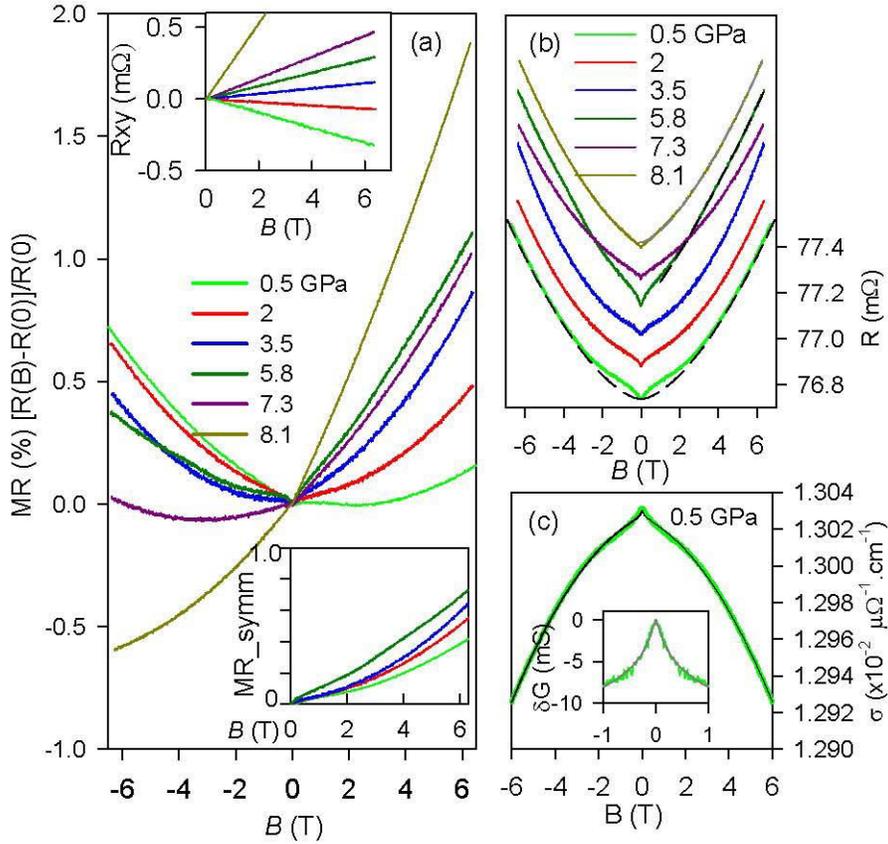

**Figure 5.** (a) Transverse magneto-resistance [MR=(R(B)-R(0))/R(0)] as a function of magnetic field at various high pressures (at 2.5K for pressures up to 5.8 GPa and at 8K for higher pressures). The upper inset shows the anti-symmetric part, the Hall contribution, as a function of magnetic field. The lower inset is the symmetrized MR curves at various pressures up to 5.8 GPa. (b) In-plane resistance R(B), after subtracting the Hall resistance, as a function of B field at various high pressures. The resistance scale is shown for the curve at 0.5 GPa. The curves for higher pressures are arbitrarily shifted for clear comparison of low field WAL features. Dashed line at 0.5 GPa is the power law fit of high field MR data. The low field deviation from the measured resistance is due to the WAL effect. (c) Plot of magneto-conductivity (MC) at 0.5 GPa and its fit with $\sigma_{tot}=\sigma_N+\sigma_{WAL}$, as discussed in text. Inset shows the negative MC correction due to WAL and its fit with 2D HLN equation [41].

## Supplementary Information for
Pressure-induced superconductivity in weak topological insulator BiSe

Pallavi Malavi[a], Arpita Paul[b], Achintya Bera[a], D V S Muthu[a], Kunjalata Majhi[a], P S Anil Kumar[a], Umesh V. Waghmare[b], A. K. Sood[a,*] and S. Karmakar[c,*]

[a]Department of Physics, Indian Institute of Science, Bangalore, 560012, India;
[b]Theoretical Sciences Unit, Jawaharlal Nehru Centre for Advanced Scientific Research, Bangalore 560064, India;
[c]HP&SRPD, Bhabha Atomic Research Centre, Trombay, Mumbai 400085, India

*S. Karmakar, **Email:** sdak@barc.gov.in

**This PDF file includes:**

    Supplementary text
    FiguresS1 to S17
    Tables S1 to S2
    SI References

**Supplementary Information Text**

**SI Materials and Methods.**
**Experimental.** High-quality single crystals of BiSe were synthesized via the modified Bridgman method [1]. High-pressure powder x-ray diffraction measurements at room temperature have been performed at the XPRESS beam line ($\lambda$ = 0.4957 A°) of the Elettra synchrotron. Single crystals were finely powdered and loaded in a DAC for measurements under quasi-hydrostatic pressures up to 18GPa with methanol-ethanol-water (MEW) (16:3:1) as pressure transmitting medium (PTM) and Au as x-ray pressure marker [2]. The 2D diffraction images were recorded on a Mar345 IP detector and these were converted to *I vs 2θ* diffraction profiles using the Fit2D software [3]. Rietveld refinements were performed for obtaining the high-pressure structural parameters using EXPGUI software [4]. For high-pressure Raman measurements, a freshly cleaved BiSe single crystal was placed in DAC with MEW pressure medium. HORIBA Jobin Yvon (HR-800) Raman spectrometer with single monochromator has been used with 660 nm laser excitation and a 50x objective of long working distance. Laser power (<5 mW) was held low enough to avoid heating of the sample.

The low-temperature resistance, transverse magneto-resistance, and Hall measurements for characterization on a 25 micron thick freshly cleaved bare sample of ~1 mm lateral dimension were first performed using the magneto-resistance measuring insert in a 7T magnetometer. High-pressure resistance measurements at low temperatures have been performed on a freshly cleaved 20 $\mu$m thick microcrystal of ~100 $\mu$m lateral dimensions. The resistance was measured using the standard four-probe method with 1mA current excitation and *ac* lock-in detection technique (using AC370 Lakeshore resistance ac bridge). A Stuttgart version DAC was used for measurements under quasi-hydrostatic pressures up to 20 GPa using finely powdered NaCl as the PTM. A pre-calibrated motorized gear was used for pressure generation in a continuous mode at ~0.2 GPa/min rate to study pressure variation of resistivity at room temperature. For measurements down to 1.4 K, the DAC was placed inside a KONTI-IT (Cryovac) cryostat, equipped with a homemade electromagnet coil (up to 0.5 T). A nonmagnetic Cu-Be DAC (Easylab) was prepared for high-field measurements under quasi-hydrostatic pressures up to 10 GPa and was inserted into a S700X SQUID magnetometer (Cryogenic Ltd) to study MR and Hall resistance up to 7-T field and also low field dc magnetic susceptibility with adapted cell background signal subtraction method. Pressures were measured by conventional ruby luminescence.

**Computational details.** Our first-principles calculations are within the framework of density functional theory as implemented in Quantum Espresso package [5]. The exchange correlation energy is determined using the gradient density approximation (GGA) as parameterized by Perdew, Burke and Ernzerhof [6]. The scalar and fully relativistic ultrasoft pseudo-potentials [7,8] have been used here to represent the interactions between ionic cores and valence electrons, and for inclusion of effects of spin-orbit coupling (SOC). Plane wave basis sets are truncated with energy cutoffs of 60 Ry and 480 Ry in representations of Kohn-Sham wave functions and charge density respectively. Discontinuity in occupation numbers of electronic states near Fermi level is smeared with Fermi Dirac distribution function with an energy width of 40 meV. The crystal structure at each pressure is obtained by optimizing lattice constants and atomic coordinates through minimization of enthalpy ($H = E + PV$). The scalar relativistic pseudo-potentials are used to optimize the crystal structure at each pressure. Brillouin Zone (BZ) integrations are sampled on uniform meshes with 12×12×2, 6×6×2 and 8×8×8 k points in the calculations of the trigonal, orthorhombic and cubic phases of BiSe respectively.

Phonon frequencies of all phases of BiSe are estimated using density functional perturbation theory (DFPT) as implemented in Quantum Espresso code [9]. Inter-atomic force constant matrices of *Pnma* and *P-3m1* phases on 4×4×1 q-points and of *Cmcm* and *Pm-3m* phases on 4×4×4 q-points are determined and used in Fourier interpolation to obtain dynamical matrices at arbitrary q points, which are diagonalized to obtain phonon frequencies. Vibrational properties of BiSe at each pressure are determined using the fully optimized crystal structure within the scalar relativistic description. In addition, phonon dispersion of cubic BiSe at high pressure is estimated using both scalar and fully relativistic pseudo-potentials. 24×24×6, 32×32×32, 24×24×24 and 32×32×8 meshes of k-points (electrons) and 4×4×1, 4×4×4, 4×4×1 and 4×4×1 q-points (phonons) are employed to compute electron-phonon coupling matrix elements of BiSe in *P-3m1*, *Pm-3m*, *Cmcm* and *Pnma* phases respectively. Superconducting Tc is estimated within BCS theory [10] using Mcmillan's equation [11,12],

$$[T_c = \frac{\omega_{log}}{1.2} exp\left[\frac{-1.04(1+\lambda)}{\lambda(1-0.62\mu^*)-\mu^*}\right] \qquad (1)$$

where $\lambda$ is the dimensionless electron-phonon coupling constant and $\mu^*$ is the Coulomb pseudopotential [12], taken in our calculations as 0.1. $\omega_{log}$ is analogous to Debye frequency ($\Theta_D$), and expressed as [11,12],

$$\omega_{log} = exp\left[\frac{2}{\lambda} \int \frac{\alpha^2 F(\omega) \log \omega}{\omega} d\omega\right] \qquad (2)$$

where $\omega$'s are the frequencies of phonon modes. $\alpha^2 F(\omega)$ is the Eliashberg spectral function.

The (001) surface of cubic BiSe (*Pm-3m* phase at 18 GPa) is modeled with a periodic supercell consisting of infinite slab and vacuum to determine the surface electronic states. A vacuum of 15 Å is included to maintain weak interactions between periodic images of slabs and interfaces. Symmetric (Se terminated) and asymmetric (Bi and Se terminated) slabs terminated at (001) surfaces of cubic BiSe are used in simulations. 1×1×10 and √2×√2×10 surface supercells are used to model the asymmetric and symmetric slabs respectively. Electronic properties (with SOC) of the symmetric and asymmetric slabs are calculated at the structure with optimized atomic positions and lattice parameters at 18 GPa. 10×10×1 uniform mesh of k-points is employed for sampling integrations over Brillouin zones of the surface supercells.

**Ab initio electronic calculations:**
**Ambient Pressure.** Our estimates of optimized lattice constants of trigonal BiSe (P-3m1 phase) are 4.24 Å and 23.28 Å, which are within typical errors (0.8-1.7 %) of GGA calculations with respect to their experimental values (4.206 Å and 22.892 Å) [Table S1]. Electronic structure of trigonal BiSe (P-3m1 phase) reveals that it is metallic in nature when effects of SOC are not included in calculation (see Fig. S1a). Valence bands (up to 3 eV below the Fermi level) are contributed by Se 4p and Bi 6p orbitals (Bi bilayer). Conduction band comprises of 6p states of Bi in the $Bi_2Se_3$ quintuple layer. Electronic structure of BiSe contains a flat band at 150 meV above Fermi level along Γ-A direction and this has been predicted earlier with first-principles calculations [1]. This flat band is contributed by 6p states of Bi atoms in the bilayer (Fig. S1b), and it indicates that contribution to the conduction band is reversed only along Γ-A direction with respect to the contribution to the rest of the conduction band (6p states of Bi atoms in the $Bi_2Se_3$ quintuple layers). SOC splits this doubly degenerate flat band (see Fig. S1c) and opens a gap of about 11 meV at all k-vectors in the Brillouin zone. Our estimated band gap is slightly underestimated with respect to its earlier theoretical estimate (42 meV) which need all electron calculations [1]. Band gap of BiSe is indirect since the conduction band minimum occurs along Γ-K direction and valence band maximum along A-L direction. BiSe is invariant under time reversal symmetry and is centrosymmetric. Hence, $Z_2$ invariant of the electronic structure determines its topological character. Our theoretical analysis based on Fu and Kane's method [13] reveals that BiSe ($P\bar{3}m1$ phase) is a weak topological insulator ($Z_2$ invariants: (0;001)), and is in agreement with earlier theoretical work [1].

**Surface State calculations.** We have determined electronic structures of symmetric (Se-terminated) and asymmetric slab (Bi/Se termination) models having 10 unit layers of thickness for the (001) surface of cubic BiSe (Pm-3m phase) at 18 GPa to investigate surface states.
Our calculated macroscopic and planar average electrostatic potentials of the symmetric and asymmetric slabs are shown in Figs. S2b, S3b. Macroscopic average is obtained by averaging planar average electrostatic potential over distances of one unit cell along the z direction. The electrostatic potential in vacuum away from the asymmetric slab has a nonzero slope (see Fig. S2b), and it suggests that the system is polar in nature and there is an electric field of $-1\times10^9$ V/m inside the vacuum. The electric field inside the asymmetric slab is negligible ($-2\times10^6$ V/m). In symmetric slab model, the magnitude of electric field inside vacuum ($-4\times10^6$ V/m) is negligibly small compared to that in the asymmetric slab model.

Electronic structure of the asymmetric slab of cubic BiSe phase (see Figs. S2c,e) exhibits metallic character. Three doubly degenerate electronic states near Fermi level emerge at M point (Bands 1, 2 and 3, compared to that of bulk) are the surface states. The doubly degenerate electronic state at M point and 84 meV below the Fermi level (Band 1, see Figs. S2c,d) is contributed by Se p states near the surface. On the other hand, the doubly degenerate electronic state at 15 meV below Fermi level (at M point) is comprised of p states of all Bi and Se atoms (Band 2). Another doubly degenerate electronic state at 72 meV above the Fermi level (Band 3) is contributed by p states of Bi and Se atoms residing in the middle of the slab. Dispersions of bands 1 and 3 are quadratic near M point. Electronic structure of asymmetric slab exhibits Dirac like linear band crossing near M point at 10 meV below the Fermi level (Band 2).
Electronic structure of the symmetric slab (see Figs. S3c,e) also exhibits surface states (marked as bands A, B, C and D) near Fermi level at Γ and M points as compared to bulk. Three doubly degenerate electronic states at Γ point and 27 meV (Band A), 116 meV (Band B) and 226 meV (Band D) above Fermi level are contributed by p states of all Bi and Se atoms (see Figs. S3d,e ). On the other hand, band C at Γ point comprises of p states of Bi and Se atoms near surface. Electronic structure of symmetric slab also shows Dirac like feature near Γ point (linear band crossing along Γ - X direction), similar to asymmetric slab. These findings highlight the co-existence of nontrivial surface states (linear band crossing) and superconductivity in cubic BiSe at high pressure.

**X-ray Diffraction.**

High pressure XRD studies show that BiSe undergoes a phase transformation sequence Trigonal (P-3m1) → Orthorhombic (simultaneous presence of *Cmcm* and *Pnma* phases) → Cubic (CsCl-type, *Pm-3m*) structure at ~8 GPa and ~13 GPa respectively, with mixed phases persisting over broad pressure ranges. Figure S4a shows the comparison of the experimental pattern at 7.8 GPa with the simulated patterns (using PowderCell software) of three BiSe phases (*P-3m1, Cmcm* and *Pnma*). Figure S4b shows the Le-Bail profile fitting of the XRD pattern at 7.8 GPa using the above combination of phases. Similar analysis of the XRD patterns at 11.1 and 13.1 GPa are shown in Figs.S4c,d and Figs. S4e,f respectively. At 11.1 GPa, simultaneous presence of orthorhombic phases (*Cmcm* and *Pnma*) are noticed. At 13.1 GPa, CsCl-type cubic phase emerges in presence of the above two phases. With further increasing pressure the fraction of the orthorhombic phases decreases while the CsCl-type BiSe phase grows rapidly, as apparent from the growing intensity of the CsCl-Bragg peak at $2\theta=11.5°$.

Structural parameters of the various phases of BiSe at different pressures are listed in Table S1. Structural evolution in different phases of BiSe is also shown in Fig. S5. The triogonal phase (*P-3m1*) is found stable up to ~7.5 GPa showing a systematic P-V plots. BM-EOS fit gives B=38.7 GPa and B'=5.4. The c-axis compressibility is twice higher than the ab-plane compressibility due to the presence of interlayer van der Waal gap. The c/a ratio is found to decrease more gradually up to 3.3 GPa and at higher pressures this decreases slowly. A similar behavior has earlier been reported in $Bi_2Se_3$, which was claimed to be due to an iso-structural transition (resulting from electronic structural modification) at this pressure [14]. Our detailed structural analysis show that Bi-Se bond distances and the Se-Bi-Se bond angles show monotonic pressure dependences, without displaying any signature of structural anomaly at this pressure.

To verify whether trigonal structured BiSe (in WTI state) undergoes any iso-structural transition (of electronic origin) which may change the compressibility at ~3 GPa without showing volume discontinuity, we have performed a more accurate analysis of the volume vs. pressure up to 9 GPa, as shown in Fig. S5f. We have performed a linearization of the BM-EOS with a plot of the reduced pressure vs the Eulerain strain (as in the case of $Bi_2Se_3$ and $Bi_2Te_3$, [15, 16]):

$$H = B_0 + \frac{2}{3}B_0(B' - 4)f_E \qquad (1)$$

where $H = \frac{P}{3.f_E(1+2f_E)^{5/2}}$ is the reduced pressure and $f_E = \frac{1}{2}\left[\left(\frac{V_0}{V}\right)^{\frac{2}{3}} - 1\right]$ is the Eulerian strain. As the data cannot be fit with a single linear fit, an iso-structural transition is clearly identified at ~3 GPa with sudden change in the compressibility at this pressure.

**Raman scattering.**

Raman spectra of BiSe at various high pressures (up to ~23 GPa) are shown in Fig. S6. At low pressures six Raman modes are observed with mode frequencies marked M1 to M6, as determined by Lorentzian fit of the spectra. While the Raman spectra remain mostly unchanged up to 6.8 GPa, the drastic change in spectra above 7.3 GPa can be associated with the structural transition into the orthorhombic structure. The significant broadening of the Raman spectra may also be attributed to the reconstructive type phase transition of BiSe. Also Raman being a local probe, the spectral broadening might be due to enhanced structural disorder of the high pressure phases, as seen in XRD data. The pressure variation of the mode frequencies in the low pressure trigonal phase are shown in Fig. S7. Although all the modes stiffen with increasing pressure, suggesting trigonal phase stability up to 6.8 GPa, noticeable slope changes in all the modes near 3 GPa (without emergence of any new peaks) can be associated with the iso-structural transition, corroborating with our XRD results. The results are also in agreement with the pressure dependencies of calculated phonon frequencies (discussed below).

**Phonon calculations.**

Zone-centre phonon frequencies of BiSe in all phases are estimated using density functional perturbation theory (DFPT) as implemented in Quantum Espresso code [9]. The trigonal unit cell of BiSe (P-3m1 phase) contains six formula units. Phonon modes at $\Gamma$ point in the P-3m1 phase can be decomposed as $\Gamma = 6A_{1g} + 6E_g + 6A_{2u} + 6E_u$. $A_{1g}$ and $E_g$ symmetric modes are Raman active whereas $A_{2u}$ and $E_u$ symmetric modes are infrared active. Three acoustic modes are labeled with $A_{2u} + E_u$. We determine the pressure dependence of Raman active mode frequencies in the P-3m1 phase (see Fig. S8). The iso-structural transition at 4.5 GPa (as discussed in pressure dependence of lattice parameters in Fig. S5 and shift of conduction band minimum in Fig. 2c-d) is also evident in change in pressure dependence of Raman active modes in the P-3m1 phase near 4.5 GPa, in agreement with our high pressure Raman scattering measurements (Fig. S7). Both $A_{1g}$ and $E_g$ symmetric modes are sensitive to pressure. Out of the twelve $A_{1g}$ and $E_g$ modes, $E_g$ mode involving in-plane displacements of Bi atoms in the Bi bilayers with frequency 94 cm$^{-1}$ (at 0 GPa) is most sensitive (slope 3.7-5.8 cm$^{-1}$/GPa) to hydrostatic pressure.

We also determined the pressure dependence of phonon frequencies (at $\Gamma$ point) in the Cmcm phase of BiSe which is stable in the range of pressure from 11 GPa to 15 GPa. The orthorhombic primitive unit cell (Cmcm phase) contains two formula units and the phonon modes are decomposed as $\Gamma = 2A_g + 2B_{1g} + 2B_{2g} + 2B_{1u} + 2B_{2u} + 2B_{3u}$. $A_g$, $B_{1g}$ and $B_{2g}$ are Raman active modes whereas $B_{1u}$, $B_{2u}$ and $B_{3u}$ are infrared active modes. Phonon modes involving Se atoms with Ag ($\omega$=180 cm$^{-1}$ at 12 GPa, slope: 2.5 cm$^{-1}$/GPa) and $B_{2g}$ ($\omega$ =141 cm$^{-1}$ at 12 GPa, slope: 2 cm$^{-1}$/GPa) symmetries are most sensitive to pressure. In the Pm-3m phase, there is only one triply degenerate infrared active optical mode of $T_{1u}$ symmetry. The observed Raman modes of the low pressure trigonal phase and the calculated Phonon frequencies in various high pressure phases of BiSe have been tabulated in Table S2.

**Superconductivity upon release of pressure.**

Upon complete release of pressure from 19.5 GPa, the R(T) behavior of BiSe crystal remains in the unconventional metallic state (dR/dT<0) (see Fig S13). More interestingly, superconducting resistance drop is observed below 2K. The resistance drops more with lowering excitation current. At the lowest temperature, as the magnetic field of 0.5 T is applied, the resistance drop smears out and the system reaches to normal metallic state. This indicates that Tc systematically decreases upon increasing field.

**Effect of SOC on EPC.**

We first calculated the vibrational spectra of BiSe in the Cmcm and Pm-3m phases at various pressures without SOC. Hydrostatic pressure results in stiffening of phonon spectra in both phases (see Fig. 4a,b) and so the electron-phonon coupling (EPC) constant λ decreases monotonically with increasing pressure (Fig. 4c).

Next, we study the effect of SOC on the electron-phonon coupling constant λ and superconducting Tc in the cubic Pm-3m phase. Spin-orbit coupling stiffens the phonon modes of BiSe at high symmetry points (see Fig. 4b). Our finding of mode stiffening is in contradiction with the SOC induced mode softening in α-Po [17] and superconducting Pb [18]. Earlier calculations on Pb and α-Po considered optimized lattice constants with SOC whereas we have not optimized lattice constants with SOC. Weights of the peaks in the Eliashberg spectral function (see Fig. 4b) are enhanced after inclusion of SOC similar to the findings in Pb [18]. The Eliashberg spectral function is expressed as [18,19],

$$\alpha^2 F(\omega) = \frac{1}{2\pi\hbar N(E_F)} \sum_{q\nu} \frac{\gamma_{q\nu}}{\omega_{q\nu}} \delta(\omega - \omega_{q\nu}) \qquad (2)$$

where $N(E_F)$ is the electronic density of states at the Fermi level and $\omega_{q\nu}$ is the frequency of the $\nu^{th}$ phonon mode at wave vector q. $\gamma_{q\nu}$ is the phonon linewidth arising from the electron-phonon interaction, and is expressed as [18,19],

$$\gamma_{q\nu} = 2\pi\omega_{q\nu} \sum_k \sum_{ij} \left|g_{k+q}^{q\nu,ij}\right|^2 \times \delta(\epsilon_{k+q,i} - \epsilon_F)\delta(\epsilon_{k,j} - \epsilon_F) \qquad (3)$$

Where $g_{k+q}^{q\nu,ij}$ 's are the electron-phonon coupling matrix elements and $\varepsilon_k$'s are the electronic band energies. $\gamma_{q\nu}$ is shown in Fig. S14 of phonons along high symmetry directions show that inclusion of SOC leads to enhancement of $\gamma_{q\nu}$ over the large part of the Brillouin zone. $\gamma_{q\nu}$ increases almost four times at Γ point (optical phonon mode) in presence of SOC. The electron-phonon coupling constant λ is related to the Eliashberg spectral function and phonon linewidth $\gamma_{q\nu}$ by [6,7] $\lambda = 2\int_0^\infty \frac{\alpha^2 F(\omega)}{\omega} = \frac{1}{\pi\hbar N(E_F)} \sum_{q\nu} \frac{\gamma_{q\nu}}{\omega_{q\nu}^2}$ (4)

As an effect of SOC, frequencies of phonons at high symmetry points increase by 6-10 cm$^{-1}$ (18 GPa) after inclusion of SOC. Additionally SOC increases $N(E_F)$ from 5.41 states/eV (without SOC) to 6.0 states/eV at 18 GPa. SOC may cause reduction in λ as $\omega_{q\nu}$ and $N(E_F)$ are inversely proportional to λ (see Eq. 4). In contrast, SOC induced significant enhancement of $\gamma_{q\nu}$ over the BZ increases λ from 0.74 (without SOC) to 0.84. This eventually leads to an enhancement of superconducting Tc from 4.9 K (without SOC) to 6.7 K (with SOC) at 18 GPa (See Fig. 4c-d). Enhancement of superconducting transition temperature emerging from SOC has been also predicted earlier in Pb and α-Pu [17,18].

**Trigonal Phase In-plane Magneto-resistance.**

In-plane (*ab*-plane) magneto-resistances in the trigonal phase were first measured on a bare single crystal BiSe flakes of 25 μm thickness.in the van der Pauw geometry (as has also been used in DAC-based high pressure measurements). Transverse Magnetoresistances are measured in standard four probe technique using *ac* lock-in technique. See Fig S15. As expected in the conventional metallic state, high field (up to 7 T) MR follows quadratic field (~$B^2$) dependence resulting from the classical cyclotron orbital motion of the carriers. At low field (below 1 T), large deviation occurs due to quantum correction to the Drude conductivity due to strong spin orbit coupling (SOC)-driven weak anti-localization (WAL) effect.

To understand the WAL contribution in the observed MR in our Bulk crystal (25 μm thick), we have studied MR under varying magnetic field orientation with respect to the *ab*-plane normal (θ= angle between the c-axis and the magnetic field direction). With B perpendicular to the *ab*-plane (θ=0°), the WAL cusp magnitude in MR is maximum (as can be clearly observed by extrapolating the quadratic field dependent curve of high field MR data (Fig S15). With increasing θ angle, the WAL part magnitude reduces systematically. With B applied along the *ab*-plane (θ=90°, the green curve) WAL feature is highly reduced, but a small amount of cusp feature still present. In an ideal 2D transport (in thin film), WAL is 2D like with magnitude ∝ *B.cosθ*. As our crystal is bulk, there is a possibility of 3D like WAL contribution. Also θ deviation from 90° by small angle (due to slight misalignment) can lead to this non vanishing WAL. In any case, a large part of WAL is originating from 2D transport nature. In our bulk crystal, 2D transport nature is suggestive of many coherent 2D channels (due to flake thickness being much larger than the phase decoherence length), as also supported by obtained large value of the α parameter (~75-125) in our data fit with the HLN equation [20].

High pressure magneto-resistance has been measured up to 8.1 GPa. The WAL-related enhanced MR can be identified by the extrapolated curve of the high field quadratic B fit. As can be seen in Fig. S16, signature of WAL (the deviation from the classical MR curve) is retained throughout the entire pressure range. Note that the measurements at 7.1 and 8.1 GPa were performed at higher temperature (~8K) above the SC Tc. As a result, the WAL looks faded. Also a slightly enhanced WAL at 5.8 GPa and a quasi-linear MR at higher field is due to the fact that at this P, SC emerges resulting in a small resistance drop at 2.5K. Analyses of the WAL feature at various pressures have been performed by calculating the negative magnetoconductivity (MC) in the low field regime, shown in Fig. S16b. Further, fitting of the MC data with the standard HLN equation for 2D transport [20] (in the limit of low carrier mobility) gives the values of phase coherence length Lφ=250-275 nm with α=75-125 for pressures up to 5.8 GPa.

-------------------------------------------

**Figures.**

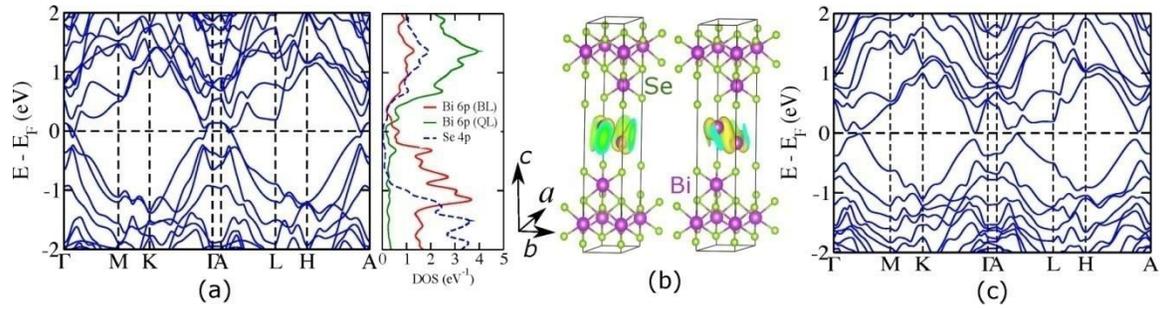

Fig. S1. Electronic structure and orbital resolved density of states of BiSe(P-3m1 phase) at P=0 GPa calculated (a) without spin-orbit coupling and (c) with including spin-orbit coupling. BL and QL in (a) represent Bi bilayer and $Bi_2Se_3$ quintuple layer respectively. (b) Iso-surfaces of charge densities of doubly degenerate electronic states at Γ (150 meV above Fermi level, conduction band minima) corresponding to the flat band along Γ-A direction in (a).

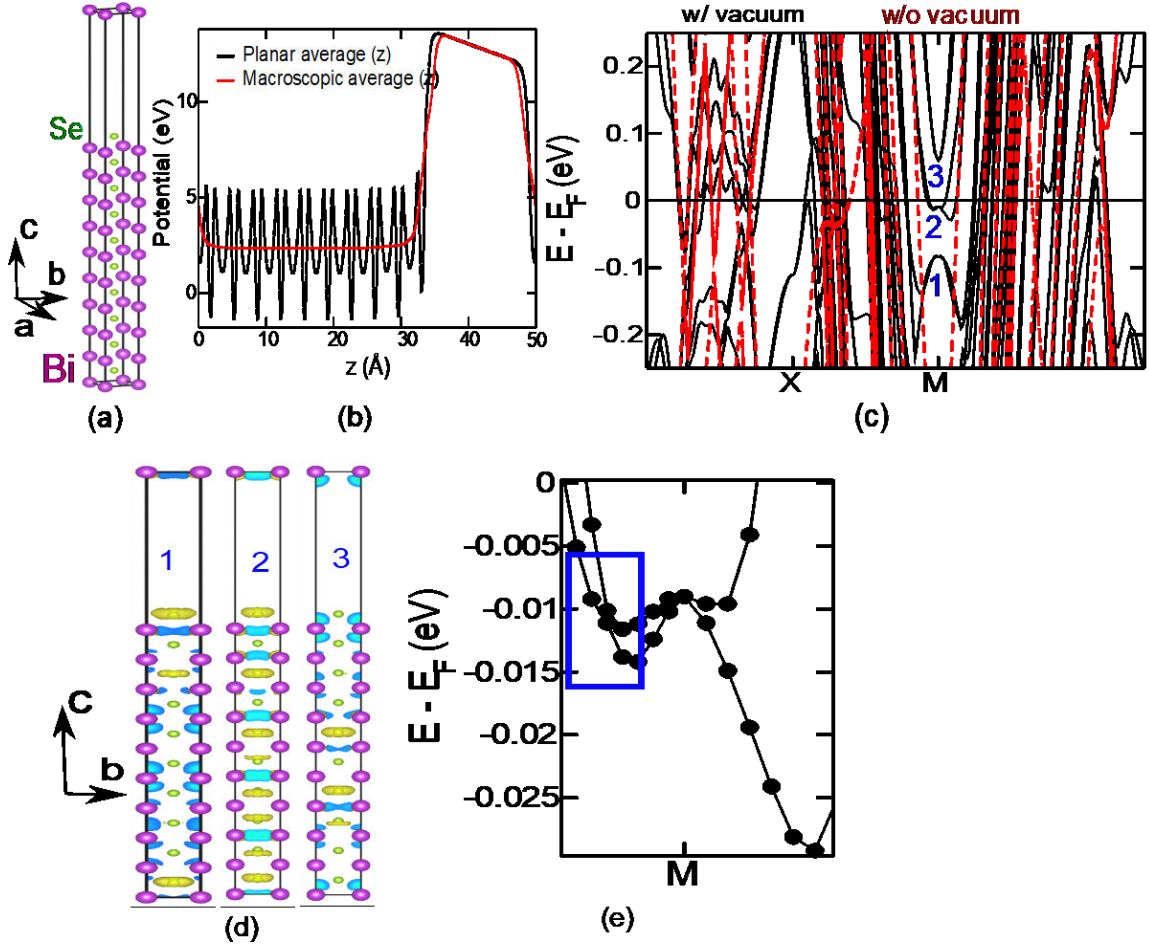

Fig. S2. (a) Structure of asymmetric slab of cubic BiSe (1×1×10 supercell with 15 Å vacuum along c direction), (b) variation in electrostatic potential along the direction perpendicular to (001) surface of cubic BiSe, and (c) comparison of electronic structures (calculated with SOC) of the asymmetric slab (1×1×10 supercell with 15 Å vacuum) and supercell (1×1×10 supercell without vacuum, i.e. bulk) of BiSe. Three doubly degenerate electronic states at M point (near Fermi level) in (c) denoted as 1 (84 meV below Fermi level), 2 (1.5 meV below Fermi level) and 3 (72 meV above Fermi level) are surface states. (d) Iso-surfaces of charge densities of three doubly degenerate electronic states (states 1, 2 and 3 in (c)) at M point. (e) Dispersion of band 2 (1.5 meV below Fermi level at M point) in (c). The blue rectangle in (e) highlights Dirac like linear band crossing along X-M direction in the electronic structure of asymmetric slab.

.

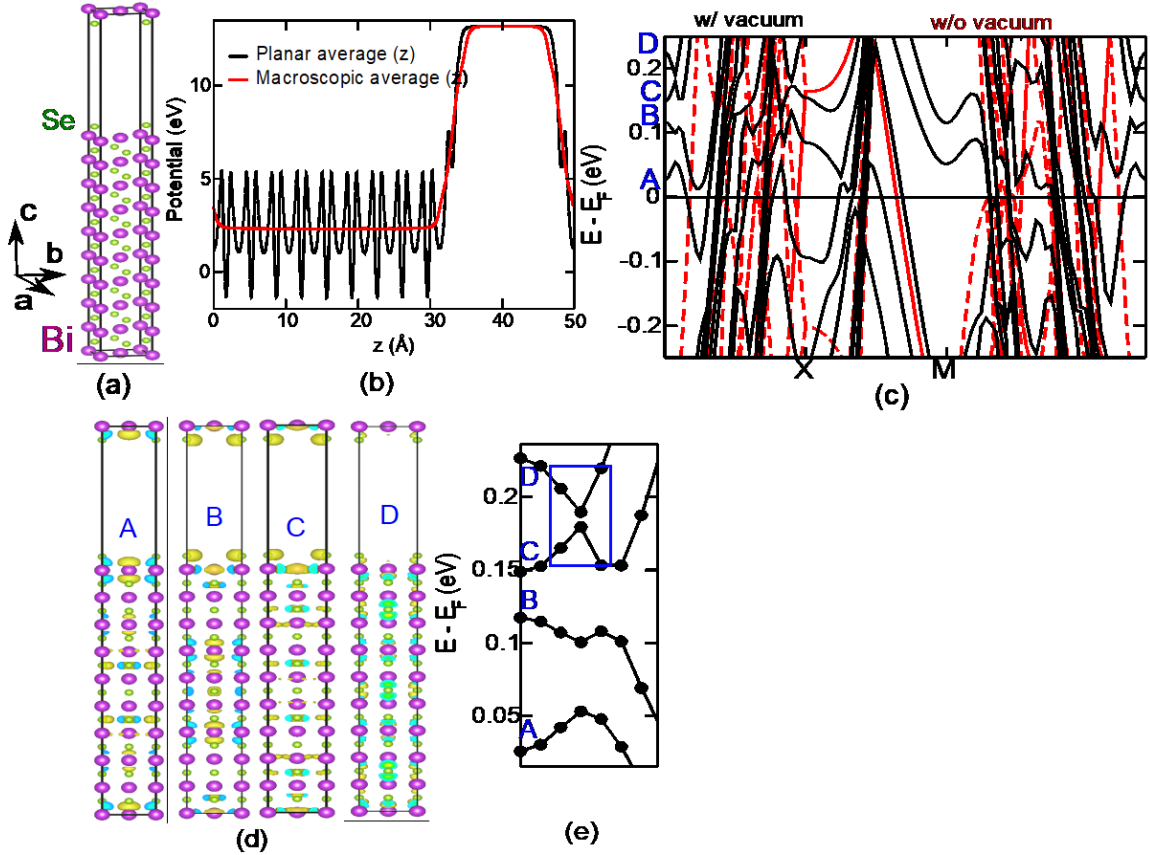

Fig. S3. (a) Structure of the symmetric slab of cubic BiSe (√2× √2× 10 supercell with 15 Å vacuum along c direction), (b) variation in electrostatic potential along the direction perpendicular to (001) surface of cubic BiSe, and (c) comparison of electronic structures (calculated with SOC) of symmetric slab (√2× √2× 10 supercell with 15 Å vacuum) and supercell (√2× √2× 10 supercell without vacuum, i.e. bulk) of BiSe. Four doubly degenerate electronic states at $\Gamma$ point in (c) marked as A (27 meV above Fermi level), B (116 meV above Fermi level), C (150 meV above Fermi level) and D (226 meV above Fermi level) are surface states. (d) Iso-surfaces of charge densities of four doubly degenerate electronic states (A, B, C and D in (c)) at $\Gamma$ point. (e) Dispersions of bands C and D in (c) near $\Gamma$ point. The blue rectangle in (e) highlights Dirac like linear band crossing along $\Gamma$-X direction) in the electronic structure of symmetric slab.

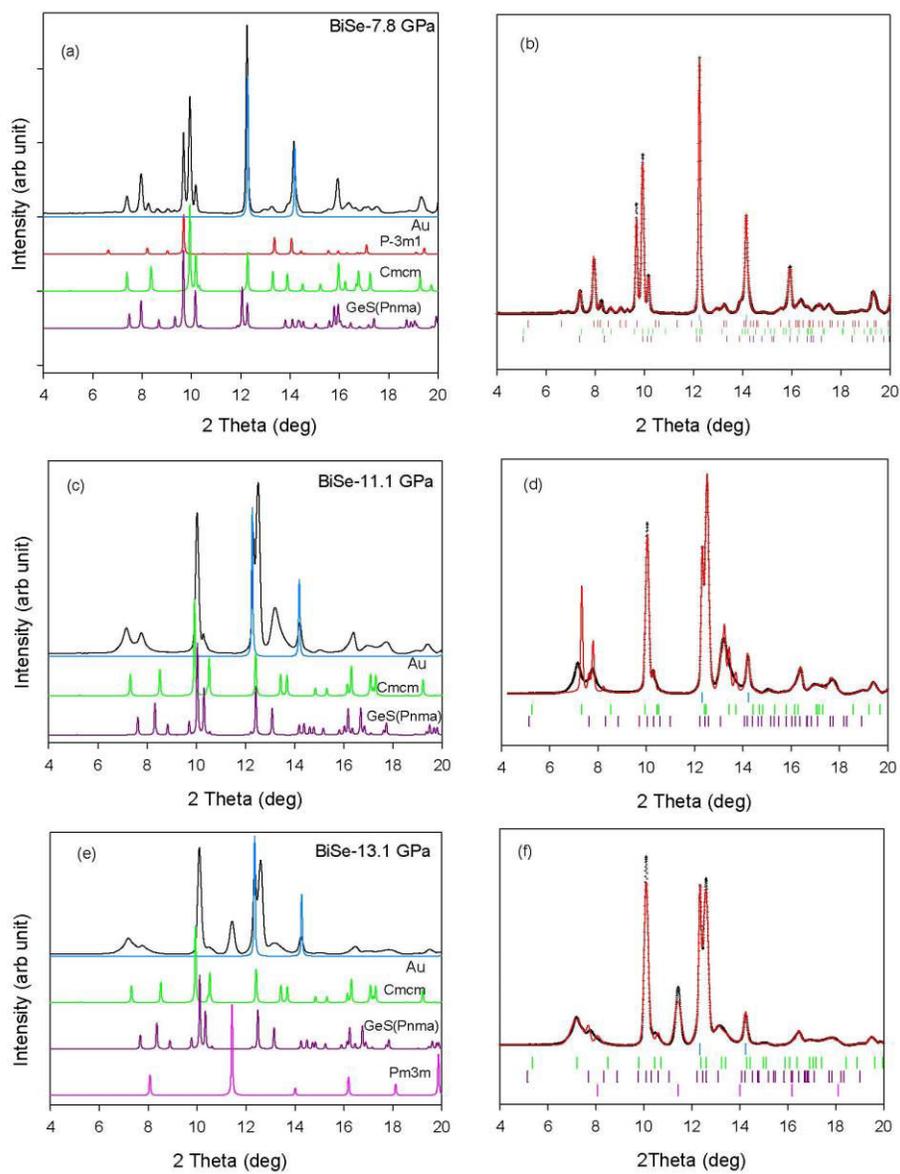

Fig. S4. (a,c,e) Comparison of experimental and simulated XRD patterns at selected pressure points. (b,d,f) Mixed phase profile fitting at those pressures.

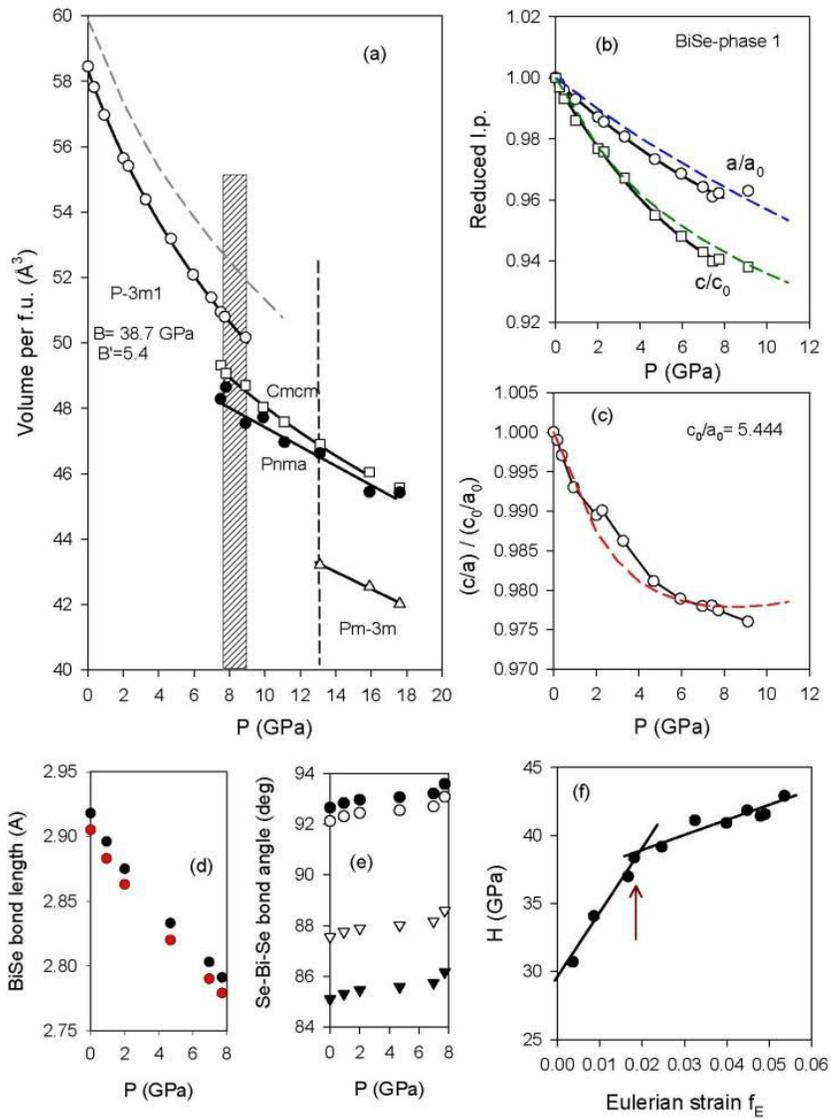

Fig. S5. (a) Volume per formula unit plotted as a function of pressure at various BiSe phases. (b) Trigonal phase reduced lattice parameters are plotted as a function of pressure (dashed lines are from theoretical calculation for the Pm-31 phase). (c) c/a raio plotted as a function of pressure. (d) Bi-Se bond length and (e) Se-Bi-Se bond angles are plotted as a function of P. (f) Reduced pressure (H), plotted as a function of the Eulerian strain (f), as discussed above

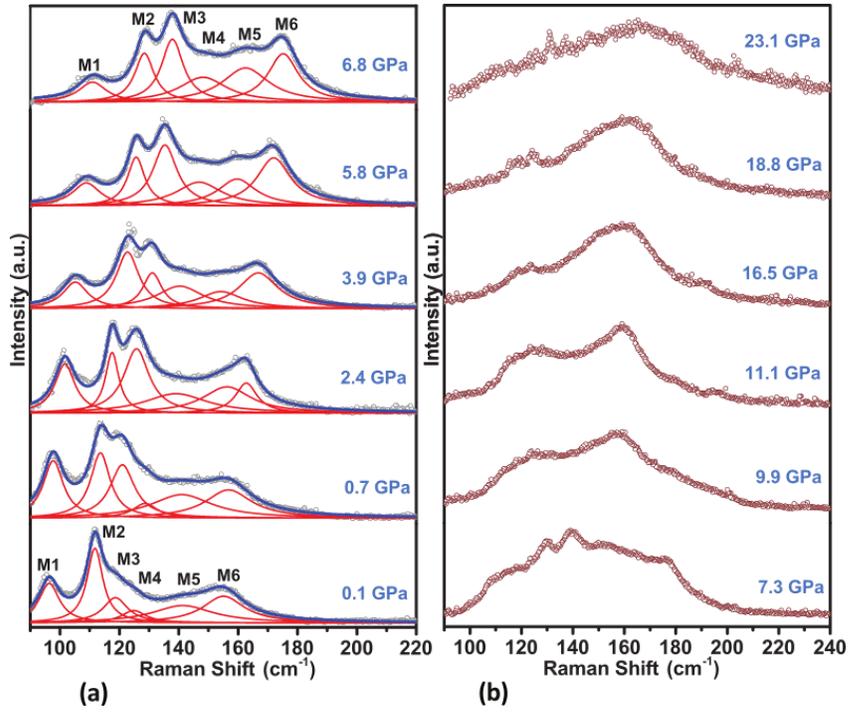

Fig. S6. Raman spectra at various high pressures. Solid lines in are the Lorentzian fits to the experimental data points. Raman modes are marked M1 to M6.

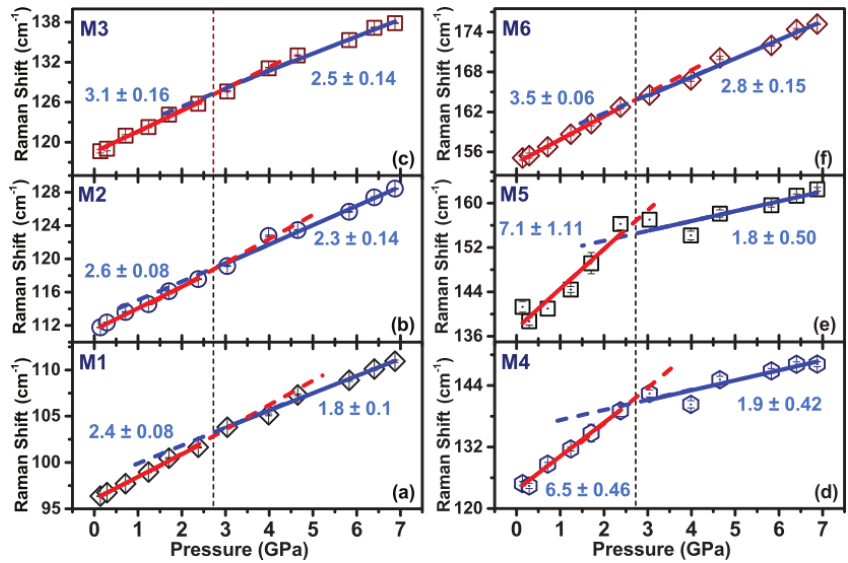

Fig. S7. Phonon frequencies (M1-M6) are plotted as a function of pressure. Solid lines are linear fits having slope changes near 3 GPa.

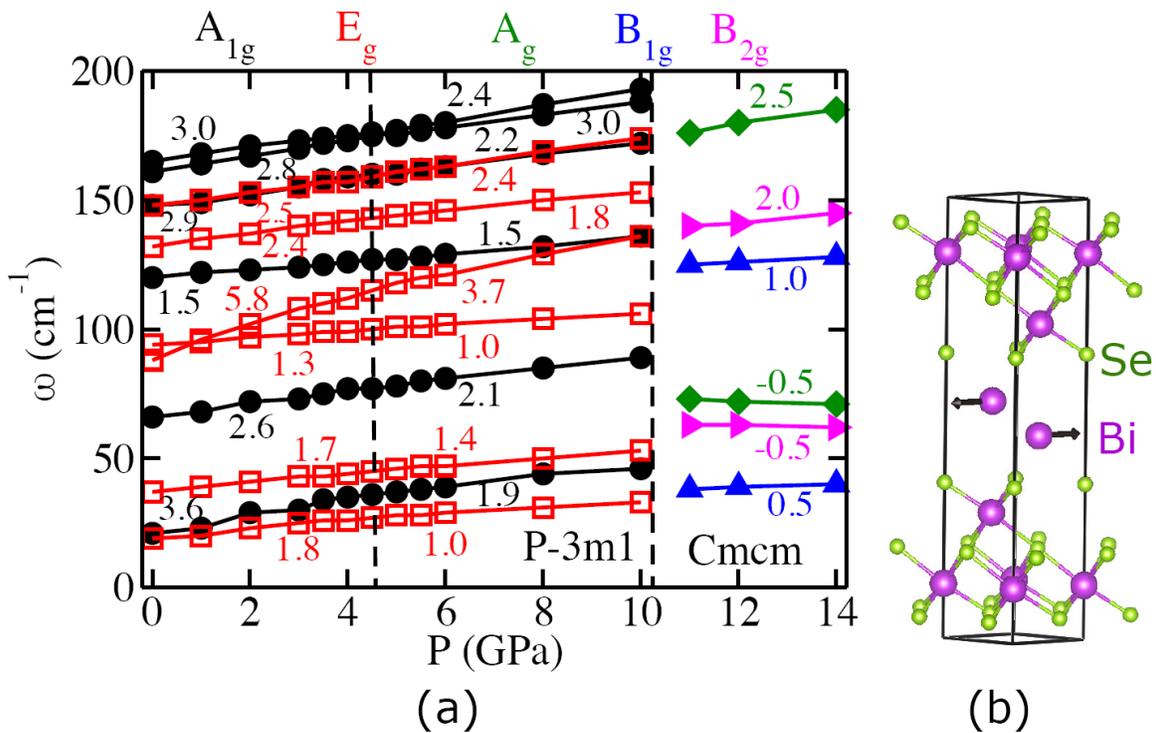

Fig. S8. (a) Pressure dependence of Raman active modes (P-3m1: A1g (black circle symbol) and Eg (red square symbol)); Cmcm: Ag (green diamond symbol), B1g (blue triangle symbol) and B2g (pink triangle symbol)) in BiSe shows the structural transitions. Vertical dashed lines at P=4.5 GPa and P=10 GPa represent critical pressures for isostructural transition (*P-3m1* space group) and structural transition from *P-3m1* phase to *Cmcm* phase respectively. (b) Eg symmetric phonon mode with frequency 94 Cm-1 at 0 GPa involves displacements of Bi atoms in the bilayer (side view).

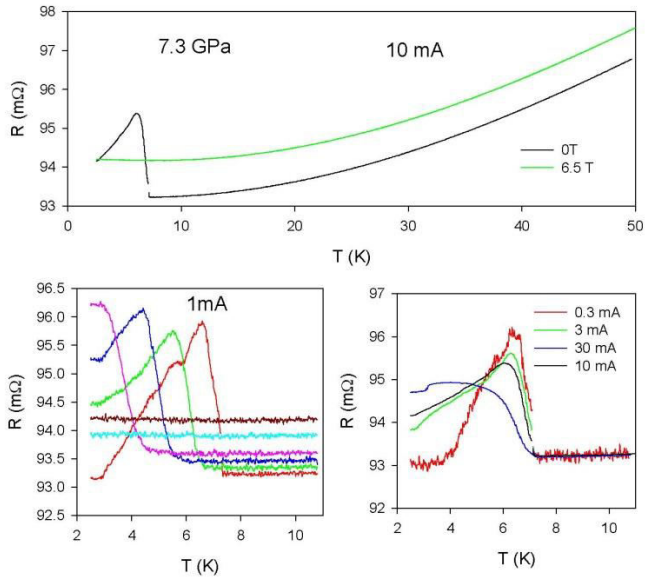

Fig. S9. (a) In-plane resistance R(T) at 7.3 GPa at 0T and 6.5 T field. (b) Field dependence of peak feature using 1mA current excitation. (c) Gradual evolution of the peak feature with increasing excitation current, at zero field.

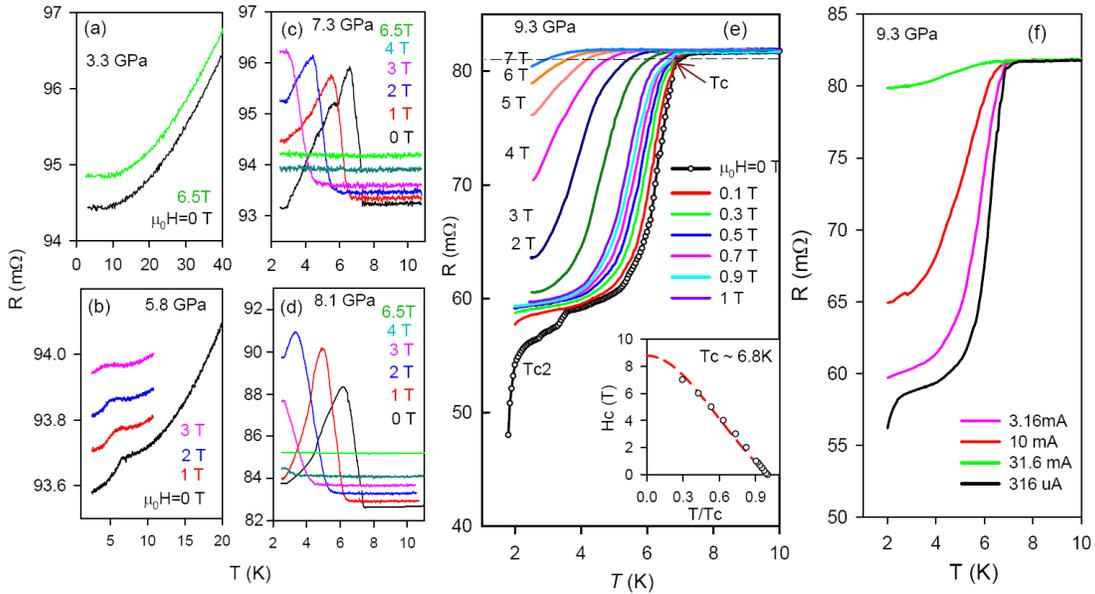

Fig. S10. R(T) plots at low temperatures under different magnetic fields, (a) at 3.3 GPa showing no SC transition, (b) at 5.8 GPa showing small resistance drop, (c,d) granularity related resistance peak formation below SC Tc at 7.3 and 8.1 GPa in the mixed phase region of the structural transition (trigonal → orthorhombic). (e) At 9.3 GPa having step-like resistance SC drops due to co-existence of two orthorhombic phases. SC Tc has been followed under magnetic field. Inset shows the Tc-H plot, the dashed line showing the Ginzburg-Landau (GL) fitted curve. (f) Variation of SC resistance drop at different current excitation.

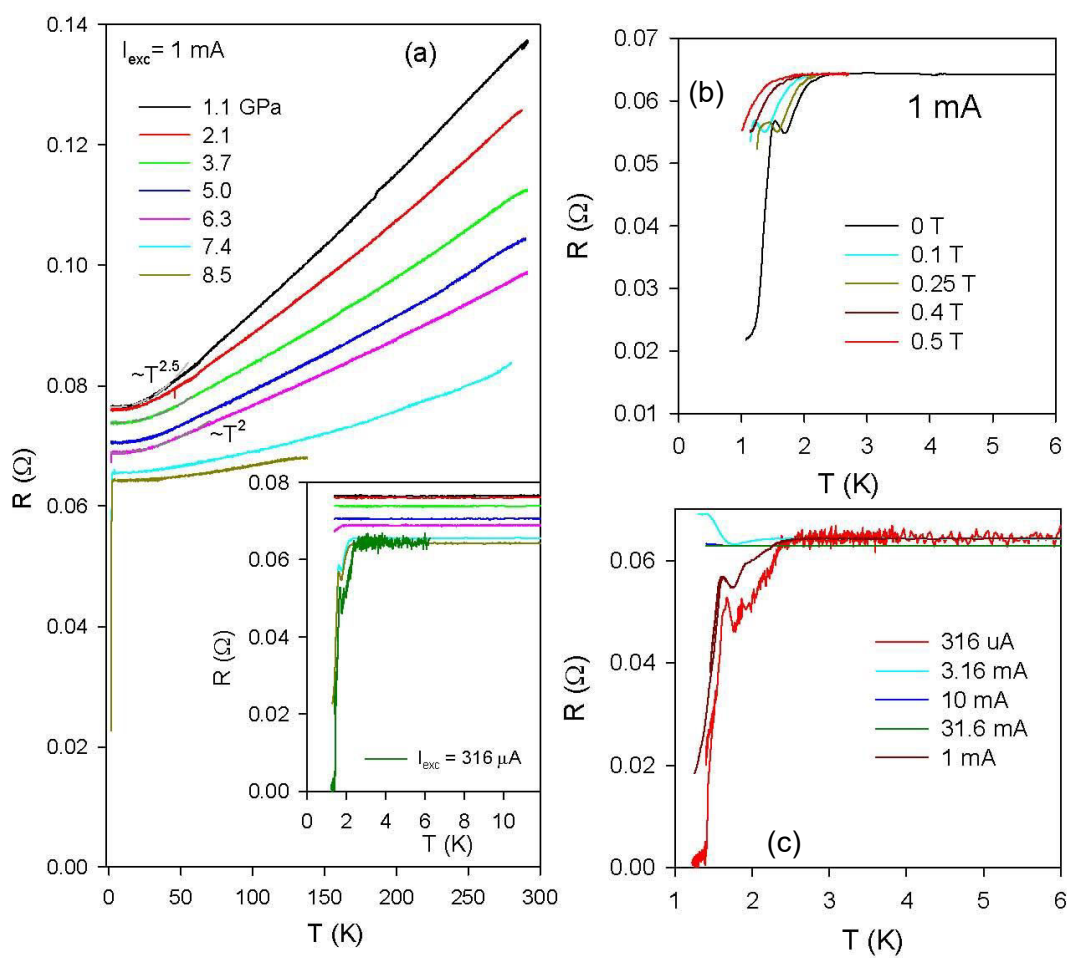

Fig. S11. (a,c) Resistance plotted as a function of temperature at various quasi-hydrostatic pressures (Run2). Inset shows expanded view of R(T) plots at low temperature near SC transition. R(T) plots at low temperature near SC transition for pressures at 8.5 GPa by varying magnetic field (b) and excitation current (c). Zero resistance is reached at 1.4K with low excitation current (316 µA).

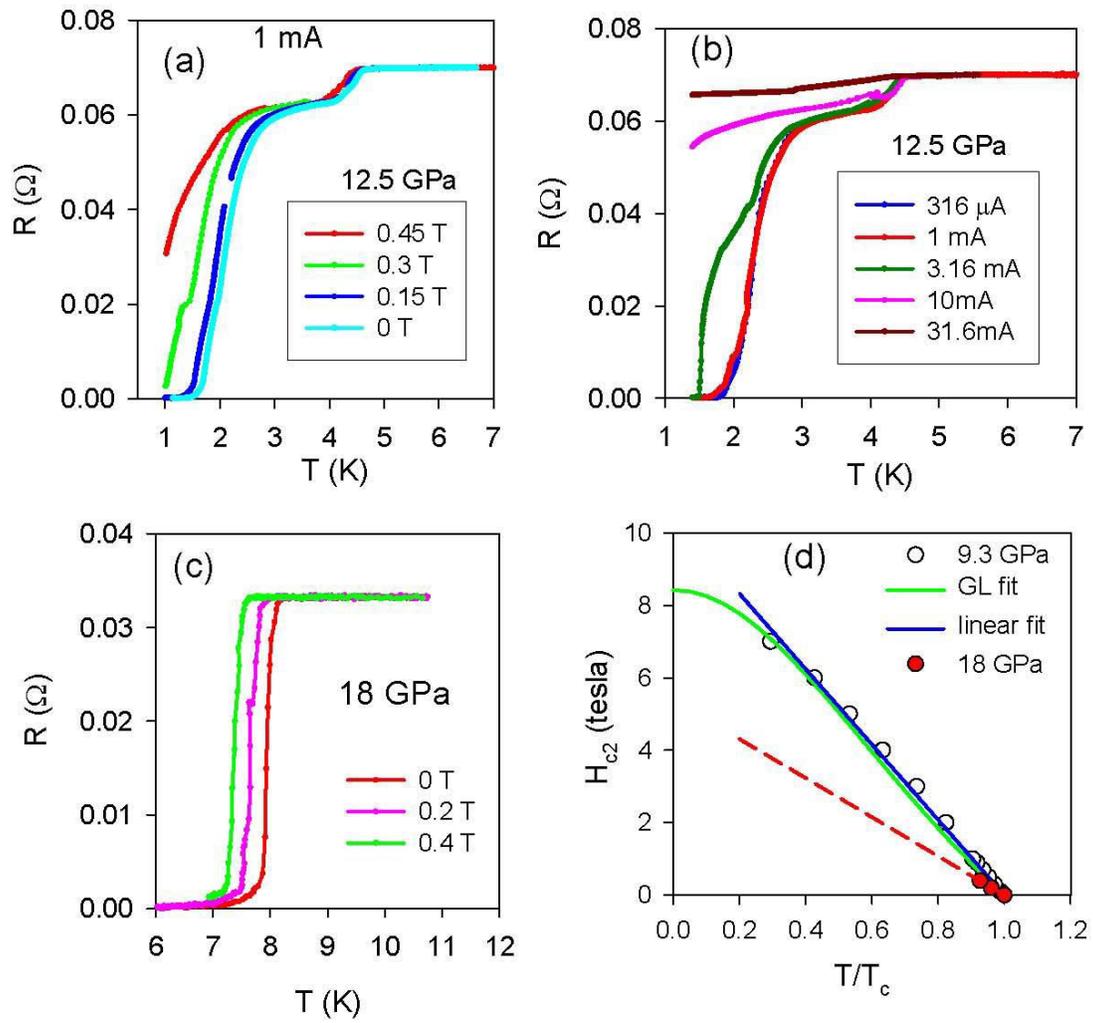

Fig. S12. (a,c) Variation of SC resistance drop under magnetic field below Tc at 12.5 GPa and 18 GPa. (b) Variation of SC resistance drop under varying current excitation at 12.5 GPa. (d) SC upper critical field has been compared for 18 GPa with that at 9.3 GPa. The estimated zero temperature critical field is significantly lower in the SC state of CsCl-type BiSe phase compared to the orthorhombic phase below 13 GPa.

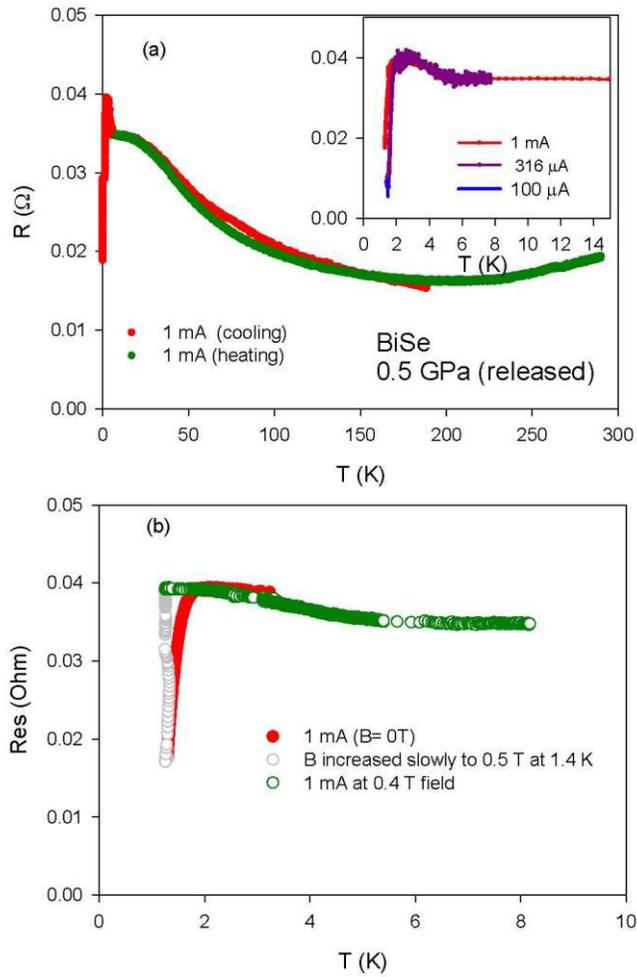

Fig. S13. (a) R(T) of BiSe upon release of pressure to ~0.5 GPa from the highest P ( ~19.5 GPa) of our study. Inset shows the increased resistance drop at lower excitation current. (b) SC resistance drop lifts systematically under application of magnetic field and normal metallic state is achieved above 1.4 K at 0.4 Tesla field.

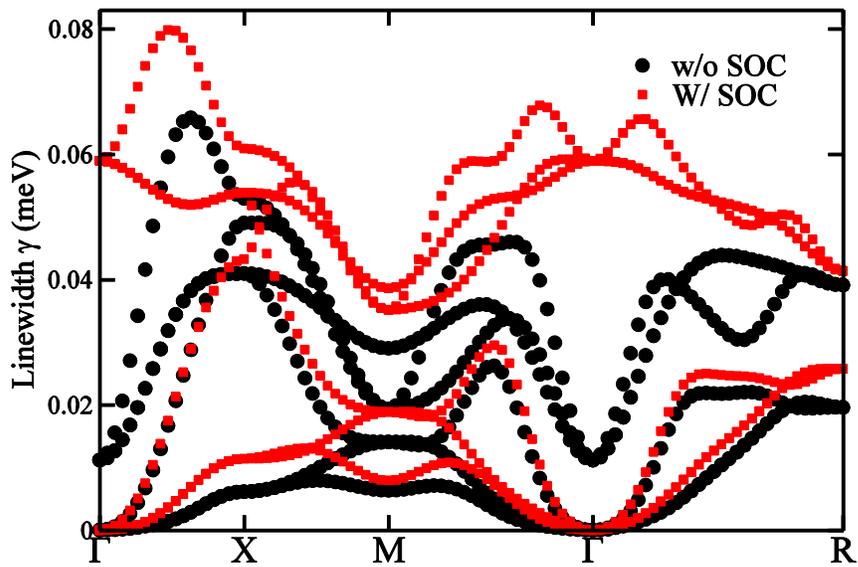

Fig. S14. Calculated phonon linewidth along high symmetry lines of BiSe in Pm-3m phase at 18 GPa. Large black (small red) points represent results obtained without (with) spin-orbit coupling.

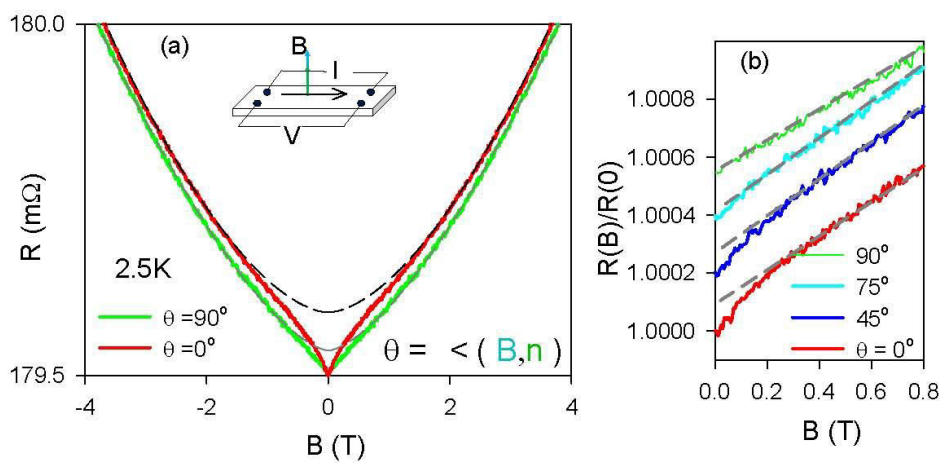

Fig. S15. In-plane (ab-plane) magneto-resistance plotted as a function of magnetic field applied in different direction with respect to the BiSe crystal ab-plane.

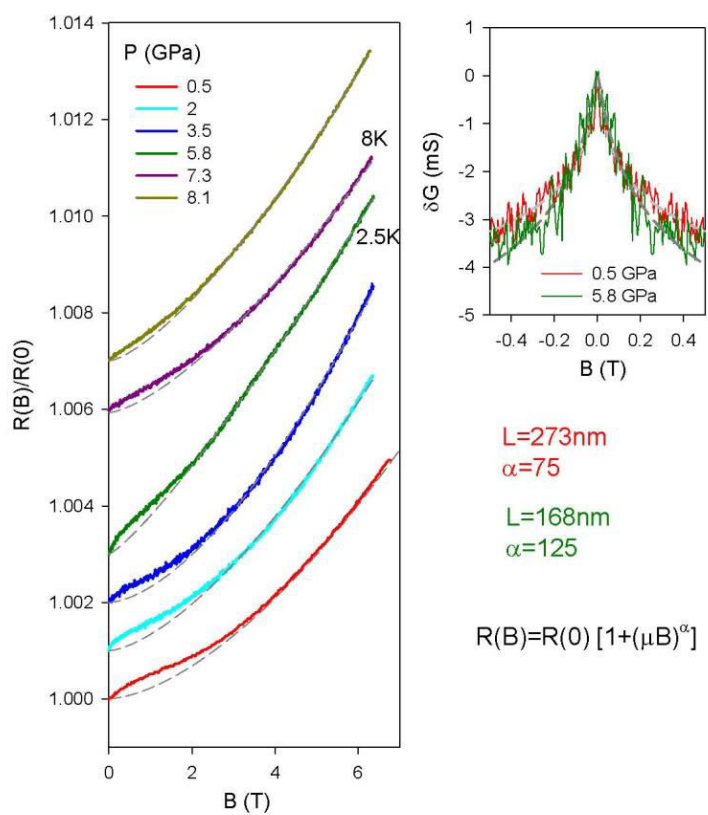

Fig. S16. (a) In-plane magneto-resistance at various pressures. The dashed lines are high field power law B fitted curve extrapolated to zero field, the deviation being originated due to WAL effect. (b) Negative magneto-conductivity (MC) at low fields, plotted for two pressures.

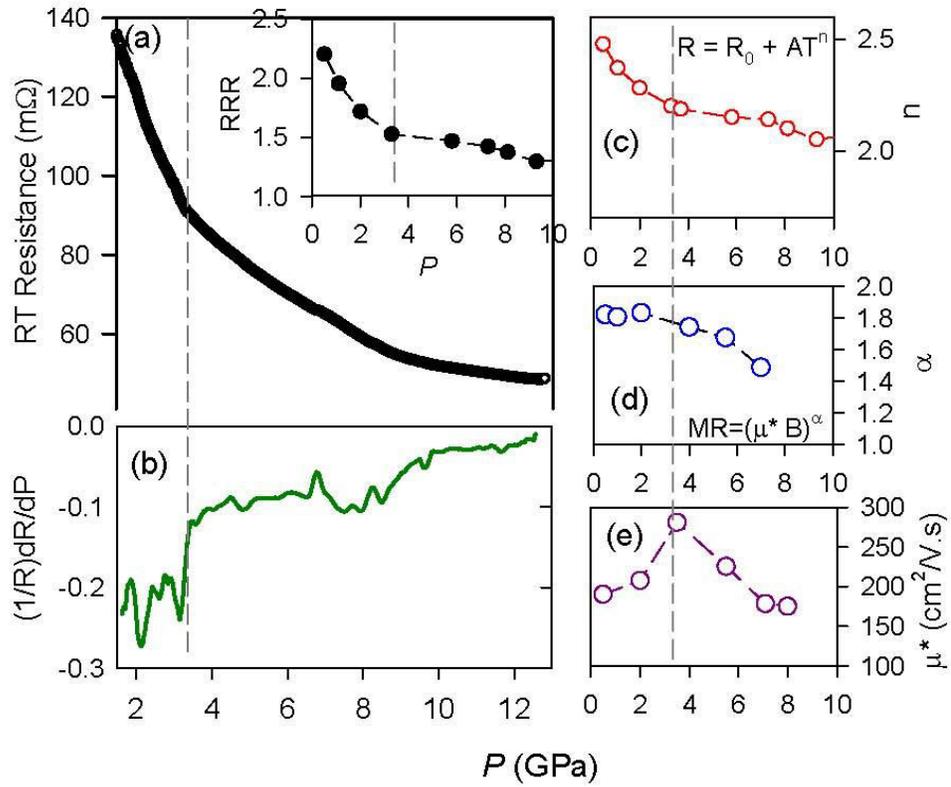

Fig. S17. (a) room temperature in-plane resistance and (b) its logarithmic derivative as a function of pressure. Inset in (a) shows the plot of RRR value as a function of pressure. Plots of (c) temperature power exponent, n of the low T normal state resistance, (d) exponent $\alpha$ and effective mobility $\mu^*$ from the MR power law fitting.

**Table S1:** Structural details as extracted from the analyses of the XRD profiles and from *ab- initio* calculations at various high pressures.

| P (GPa) | Crystal structure Space group | Lattice constants Expt | Lattice constants Theory | Atoms (sites) | x | y | z | Bulk modulus and its P-derivative |
|---|---|---|---|---|---|---|---|---|
| Ambient | Trigonal *P-3m1, z=6* | a= 4.2059 Å c= 22.8923 Å V0= 350.7 Å$^3$ | 4.24 Å 23.28 Å | Se1(2d) Bi1(2d) Bi2(2d) Se2(2d) Bi3(2c) Se3(2c) | 1/3 1/3 1/3 1/3 0 0 | 2/3 2/3 2/3 2/3 0 0 | 0.056 0.291 0.541 0.789 0.874 0.638 | $B_0$=38.7 GPa B'= 5.4 (fitting up to 8 GPa) |
| 11 GPa | Orthorhombic *Cmcm, z=4* | a= 4.285 Å b= 11.182 Å c= 4.122 Å V=197.5 Å$^3$ | 4.74 Å 10.97 Å 3.62 Å | Bi (4c) Se (4c) | 0 0 | 0.124 0.356 | 0.25 0.25 | $B_0$~101 GPa B'=5.3 (f) |
| 11 GPa | Orthorhombic *Pnma, z=4* | a= 11.1153 Å b= 4.3728 Å c= 3.9528 Å V=192.1 Å$^3$ | 11.21 Å 4.61 Å 3.69 Å | Bi (4c) Se (4c) | 0.1203 0.359 | 0.25 0.25 | 0.07 0.02 | Not calculated due to scattered data |
| 17.6 GPa | Cubic (CsCl) *Pm-3m, z=1* | a= b = c= 3.4769 Å V=42.03 Å$^3$ | 3.49 Å | Bi (1b) Se (1a) | 0.5 0 | 0.5 0 | 0.5 0 | Not calculated |

**Table S2:** Raman active phonons in BiSe phases from Raman measurements and DFT calculations

| Trigonal (P-3m1) | | Orthorhombic (Cmcm) | Cubic (Pm-3m) |
|---|---|---|---|
| Experiment (Raman modes) | Theory (at 0 GPa) | Theory (at 12 GPa) | Theory (at 18 GPa) |
| M1 ~96 cm$^{-1}$ M2 ~111 cm$^{-1}$ M3 ~118 cm$^{-1}$ M4 ~125 cm$^{-1}$ M5 ~141 cm$^{-1}$ M6 ~155 cm$^{-1}$ | Raman active modes 6A$_{1g}$ 21, 66, 119, 146, 160, 165 cm$^{-1}$ 6E$_g$ 18, 36, 86, 94, 132, 147 cm$^{-1}$ IR Active modes 5A$_{2u}$ 26, 66, 136, 153, 167 cm$^{-1}$ 5E$_u$ 22, 39, 81, 130, 145 cm$^{-1}$ Acoustic modes: A$_{2u}$ +E$_u$ | Ag (Bi and Se): 72 cm$^{-1}$ Ag (Se): 180 cm$^{-1}$ B1g (Bi): 39 cm$^{-1}$ B1g (Se): 126 cm$^{-1}$ B2g (Bi and Se): 63 cm$^{-1}$ B2g (Se): 141 cm$^{-1}$ | No Raman active mode IR active mode T$_{1u}$ 125 cm$^{-1}$ |

**SI References**